\documentclass[twocolumn,
               aps,
               prd,   
               preprintnumbers,
               numbers,
               sort&compress,
               nofootinbib,
               showpacs,
               colorlinks,
               linkcolor=blue,   
               citecolor=blue]{revtex4-1}

\usepackage{graphicx,amsmath,amssymb,bm}
\usepackage{psfrag}
\usepackage{feynmp}
\usepackage{hyperref}
\usepackage{graphicx}
\usepackage{subcaption}
\usepackage{placeins}
\usepackage{xcolor} % Enables coloured text
\usepackage{gensymb}

\def\<{\langle}
\def\>{\rangle}

\def\+{\dagger}

\def\U1A{U(1)$_{\rm A}$}

\def\ra{\rangle}
\def\la{\langle}
\def\rmd{\mathrm{d}}
\newcommand{\be}{\begin{eqnarray}}
\newcommand{\ee}{\end{eqnarray}}
\newcommand{\beq}{\begin{equation}}
\newcommand{\eeq}{\end{equation}}
\newcommand{\exclude}[1]{}
\newcommand{\jcap}{JCAP}

\newcommand{\physrep}{Physics Reports}

\newcommand{\kmps}{\,\mathrm{km}\,\mathrm{s}^{-1}}

\begin{document}

%\twocolumn[\hsize\textwidth\columnwidth\hsize\csname@twocolumnfalse\endcsname
\title{Axion Quark Nugget Dark Matter: Time Modulations and Amplifications}

\author{Xunyu Liang$^{1}$}
\email{xunyul@phas.ubc.ca}
\author{Alexander Mead$^{1,2}$}
\email{alexander.j.mead@googlemail.com}
\author{Md Shahriar Rahim Siddiqui$^{1}$}
\email{shahriar.naf07@gmail.com}
\author{Ludovic Van Waerbeke$^{1}$}
\email{waerbeke@phas.ubc.ca}
\author{Ariel Zhitnitsky$^{1}$}
\email{arz@phas.ubc.ca}
\affiliation{\\
1 Department of Physics and Astronomy, University of British Columbia, Vancouver, Canada\\
2 Institut de Ci\`encies del Cosmos, Universitat de Barcelona, Barcelona, Spain
}

%\date{\today}
\label{firstpage}

\begin{abstract}
\exclude{We advocate a new strategy in axion dark matter searches by studying the time modulations and varies correlations.
%, instead of coherence, of signals when axions present a much stronger fluctuation than the presumed model of cold dark matter halo.
We focus on a new mechanism of the axion production suggested recently in \cite{Fischer:2018niu,Lawson:2019cvy} in framework of the so-called axion quark nugget (AQN) dark matter model. We find the density of the AQN-induced axions reveal nontrivial time modulations and amplifications. In particular, the daily modulation is at least three order of magnitude stronger than in the case of conventional DM galactic axions, and there is a transient burst-like amplification of the axion density by $10^2-10^3$ for a short period of time. Numerical simulations suggest the AQN-induced axions have density in range $(10^{-6}-10^{-5})\rm\,GeV\,cm^{-3}$ with typical velocity $\langle v_a\rangle\sim0.6c$ on the Earth's surface. The present work studies the AQN-induced axions within the mass window $10^{-6}{\rm\,eV}\lesssim m_a\lesssim10^{-3}\rm\,eV$. We also comment on  the  broadband detection strategy to search for such relativistic axions by studying the daily and  annual time modulations
as well as random burst-like   amplifications.    
}

We   study  the   new mechanism of the axion production  suggested recently in \cite{Fischer:2018niu,Lawson:2019cvy}. This mechanism   is based on the so-called Axion Quark Nugget (AQN) dark matter model, which was originally invented to explain the similarity of the dark and visible cosmological matter densities.  We perform numerical simulations to evaluate the axion flux on the  Earth's surface. % based on realistic models of AQN velocity and mass distributions. 
%Unlike the conventional galactic axions, the energy flux of AQN axions is independent on the axion mass $m_a$ or initial misalignment angle $\theta_0$.  
We examine   annual and  daily modulations, which have been studied previously and are known to occur for any type of dark matter.  We also discuss a novel type of short time enhancements which are unique to the AQN model: the statistical fluctuations and burst-like amplification, both of which can drastically amplify  the axion signal, up to a factor $\sim10^2-10^3$ for a very short period of time.
The present work studies the AQN-induced axions within the mass window $10^{-6}{\rm\,eV}\lesssim m_a\lesssim10^{-3}\rm\,eV$ with typical velocities $\langle v_a\rangle\sim0.6c$. We also comment on  the  broadband detection strategy to search for such relativistic axions by studying the daily and  annual time modulations
as well as random burst-like   amplifications.    
%We find that the order of magnifications is, to some extent, reversed compared to conventional dark matter (DM) candidates such as Weakly Interacting Particles (WIMPs). 
%In particular, effect of daily modulation is comparable to annual modulation. 
%And axion signal can be drastically amplified ($\sim10^2-10^3$) if the collection time is reduced down to subsecond level, in contrast to the traditional concept of time-coherence enhancement in axion detections. %The result may also apply to other DM candidates that have hadronic interaction. 
%A number of the axion search experiments  with broad-band detection technique   may be able to detect these time dependent effects. 

\end{abstract}
\vspace{0.1in}

%Keywords:{dam , axi}

\maketitle
\section{Introduction}
\label{sec:introuction}
The Peccei-Quinn mechanism, accompanied by axions, remains the most compelling resolution of the strong charge-parity (CP) problem 
\cite[orignal papers:][]{1977PhRvD..16.1791P,1978PhRvL..40..223W,1978PhRvL..40..279W,KSVZ1,KSVZ2,DFSZ1,DFSZ2}\cite[recent reviews:][]{vanBibber:2006rb, Asztalos:2006kz,Sikivie:2008,Raffelt:2006cw,Sikivie:2009fv,Rosenberg:2015kxa,Marsh:2015xka,Graham:2015ouw,Ringwald:2016yge,Battesti:2018bgc,Irastorza:2018dyq,Safronova:2017xyt}. In this model, conventional dark-matter (DM) axions are produced either by the misalignment mechanism \cite{1983PhLB..120..127P,1983PhLB..120..133A,1983PhLB..120..137D}, when the cosmological field $\theta(t)$ oscillates and emits cold axions before it settles at a minimum, or via the decay of topological objects \cite{Chang:1998tb,2012PhRvD..85j5020H,2012PhRvD..86h9902H,Kawasaki:2014sqa,Fleury:2015aca,Gorghetto:2018myk,Klaer:2017ond}. There are some uncertainties in the estimation of the axion abundance from these two channels and we refer the reader to the original papers for discussions\footnote{\label{DM}According to recent computations \cite{Klaer:2017ond} the axion contribution to $\Omega_{\rm DM}$ as a result of decay of topological objects can saturate the observed DM density today if the axion mass is in the range $m_\mathrm{a}=(2.62\pm0.34)10^{-5} {\rm eV}$, while earlier estimates suggest that saturation occurs at a larger axion mass. There are additional uncertainties in this result, and we refer to the original studies \cite{Gorghetto:2018myk} on this matter. One should also emphasize that the computations  \cite{Chang:1998tb,2012PhRvD..85j5020H,2012PhRvD..86h9902H,Kawasaki:2014sqa,Fleury:2015aca,Gorghetto:2018myk,Klaer:2017ond} have been performed with assumption that Peccei-Quinn symmetry was broken after inflation.}. In both production mechanisms, the axions are radiated  as non-relativistic particles with typical galactic velocities $v_{\rm axion}/c\sim 10^{-3}$, and their contribution to the cosmological DM density scales as $\Omega_{\rm axion}\sim m_a^{-7/6}$. This scaling implies that the axion mass must be tuned to $m_a\simeq 10^{-5}$ eV in order to saturate  the observed cosmological DM density today. Higher axion masses will contribute very little to $\Omega_{\rm DM}$ and lower axion masses will over-close  the Universe which would be in strong conflict with cosmological data \cite{Planck2018}. Cavity-type experiments have the potential to discover these non-relativistic axions.

Note that the type of axions being discussed here are conventional Quantum Chromo-Dynamics (QCD) axions with a mass range of ($10^{-6} {\rm eV} \lesssim m_a\lesssim 10^{-3} {\rm eV}$). It should be contrasted with Ultra Light Axions (ULAs) which were suggested as another Dark Matter candidate, also called fuzzy dark matter, but it is not the kind of axions considered in this study. ULAs do not solve the strong CP problem, and their primary motivation is not rooted in QCD.  

Axions may also be produced via the Primakoff effect in stellar plasmas at high temperature \cite{Sikivie:1983ip}. These axions are ultra-relativistic; with a typical average energy of axions emitted by the Sun of $\la E\ra =4.2$ keV \cite{Andriamonje:2007ew}. Searches for Solar axions are based on helioscope instruments like CAST (CERN Axion Search Telescope) \cite{Andriamonje:2007ew}. 

Recent work \cite{Fischer:2018niu} has suggested a fundamentally novel mechanism for axion production in planets and stars, with a mechanism rooted in the so-called axion quark nugget (AQN) dark matter  model \cite{Zhitnitsky:2002qa}.
The AQN construction in many respects is 
similar to the original quark nugget model suggested by Witten \cite{Witten:1984rs} long ago (refer to \citep{Madsen:1998uh} for a review). This type of DM model   is ``cosmologically dark", not because of the weakness of their interactions, but due to their small cross-section-to-mass ratio, which scales down many observable consequences of an otherwise strongly-interacting DM candidate. 

There are two additional elements in our AQN model compared to \cite{Witten:1984rs,Madsen:1998uh}. First, there is an additional stabilization factor for the nuggets provided by the axion domain walls which   are produced in great quantity during the  QCD  transition which help to alleviate a number of problems with the original nugget model\footnote{\label{first-order}In particular, a first-order phase transition is not required as the axion domain wall plays the role of the squeezer. Another problem with \cite{Witten:1984rs,Madsen:1998uh} is that nuggets will likely evaporate on a Hubble time-scale even. For the AQN model this argument is not applicable because the vacuum-ground-state energies inside (color-superconducting phase) and outside (hadronic phase) the nugget are drastically different. Therefore, these two systems can coexist only in the presence of an external pressure, provided by the axion domain wall. This contrasts with the original model \cite{Witten:1984rs,Madsen:1998uh}, which must be stable at zero external pressure.}.  Another feature of AQNs is that nuggets can be made of {\it matter} as well as {\it antimatter} during the QCD transition. This element of the model completely changes the AQN framework \cite{Zhitnitsky:2002qa} because the DM density, $\Omega_{\rm DM}$, and the baryonic matter density, $\Omega_{\rm visible}$, automatically assume the  same order of magnitude  $\Omega_{\rm DM}\sim \Omega_{\rm visible}$ without any fine tuning.  This is because they have the same QCD origin  and are both proportional to the same fundamental dimensional parameter $\Lambda_{\rm QCD}$  which ensures that the relation $\Omega_{\rm DM}\sim \Omega_{\rm visible}$  
always holds. 

The existence of both AQN species  explains  the observed asymmetry between matter and antimatter as a result of separation of the baryon charge when some portion of the baryonic charge is hidden in the form of AQNs. Both AQNs made of matter and antimatter serve as dark matter.  It should be  contrasted with the conventional baryogenesis paradigm when extra baryons (1 part in $10^{10}$)  must be produced during the early stages of the evolution of the Universe. While the model was invented to explain the observed relation $\Omega_{\rm DM}\sim \Omega_{\rm visible}$, it may also explain a number of other observed phenomena, such as the excess of diffuse galactic emission in different frequency bands, including the 511 keV line. The  AQNs may also offer a resolution to the so-called ``Primordial Lithium Puzzle" \cite{Flambaum:2018ohm} and to the ``The Solar Corona Mystery"  \cite{Zhitnitsky:2017rop,Raza:2018gpb} and may also explain the recent EDGES observation \cite{Lawson:2018qkc}, which is in some tension with the standard cosmological model. It may also explain the DAMA/LIBRA annual modulation as argued in \cite{Zhitnitsky:2019tbh}. It is the {\it same set} of physical  parameters of the model which were used in  aforementioned phenomena that   will be adopted for the present studies. 

We refer to the original papers \cite{Liang:2016tqc,Ge:2017ttc,Ge:2017idw,Ge:2019voa} on the formation mechanism  when the   separation of baryon charge phenomenon plays the key role. The AQN framework aims at resolving  two fundamental problems at once: the nature of dark matter and the asymmetry between matter and antimatter. This is irrespective of the specific details of the model, such as the axion mass $m_a$ or misalignment angle $\theta_0$.

AQNs are composite objects made out of axion field  and quarks and gluons  in the color superconducting (CS) phase, squeezed by a domain wall (DW) shell.
It represents a cosmologicaly stable configuration state assuming  the lowest energy state for a given baryon charge. 
%We note that the  quarks and gluons contribute about $2/3$ of the total mass and can be made out of either matter or antimatter due to stability of the CS phase. 
An important point, in the context of the present study, is that the axion portion of the energy   contributes to about  1/3 of the total mass in form of the axion DW. This DW  is a    stable time-independent configuration  which  kinematically cannot convert its energy to freely propagating (time-dependent) axions. However, any time-dependent perturbation 
results in the emission of real propagating axions via DW oscillations. The axion emission happens whenever annihilation events occur  between antimatter AQNs with  surrounding material.  

In particular when AQNs cross the Earth
%, axions are emitted with typical speed $\langle v_a\rangle\simeq0.6c$ and
the corresponding axion energy density is estimated \cite{Fischer:2018niu} as:
\begin{equation}
\label{eq:rho_a intro}
\rho_a^{\rm AQN}
\sim10^{-4}\left(\frac{\Delta B}{B}\right)\rm \frac{GeV}{cm^3} ~~~~~~ [\rm AQN -induced ],
\end{equation}
where $\Delta B/{B}$ is the portion of the baryon charge being annihilated during the passage of the AQN through the Earth.
This portion  $\Delta B/{B}$ is estimated on the level $(10\%-30\%)$ depending on the size distribution of the AQNs  \cite{Lawson:2019cvy}. If the conventional galactic axions saturate the DM density $\rho_{\rm DM}\simeq0.3\,\rm GeV\,cm^{-3}$ then the estimate (\ref{eq:rho_a intro}) suggests that the AQN-induced axion density is  about four orders of magnitude smaller than the conventional galactic axions density.  
\exclude{One should comment here that   the conventional contribution is highly sensitive to $m_a$
as $\rho_{\rm DM} \sim m_a^{-7/6}$ and may saturate the DM density at $m_a\lesssim 10^{-5} {\rm eV}$, depending on additional assumptions on production mechanism. It should be contrasted with the AQN-induced density   (\ref{eq:rho_a intro}) which  is  not very sensistive to the  axion mass $m_a$ assumiong the AQNs saturate the dark matter density. }
 
 The key distinct feature between the conventional galactic axions and AQN-induced axions is the spectrum. In the former case $v_a\sim 10^{-3}c$, while for the latter, the spectrum is very broad  and demonstrates a considerable variation in the entire interval $v_a\in(0,c)$
    with average velocity $\la v_a\ra\simeq 0.6 c$ \cite{Liang:2018ecs}.
       This crucial difference in spectrum requires a different type of instruments and drastically different search strategies. A possible broadband detection strategy for the relativistic AQN-induced axions (\ref{eq:rho_a intro}) has been discussed   recently in an accompanying  paper \cite{Budker:2019zka}, where it is suggested to analyse the annual and daily modulations and different options are proposed in order to discriminate the true signals from spurious signals by using the global network.  Appendix \ref{broadband} reviews some of the ideas  advocated  in \cite{Budker:2019zka}.

The main goal of the present work is the computation of the intensity and time-dependence of all the
effects discussed in \cite{Budker:2019zka}, assuming that the broadband detection strategy will be available in the future.
  It is clear that the AQN-induced axion density \eqref{eq:rho_a intro} is very small and the signal will be strongly suppressed in conventional cavity axion search  experiments such as ADMX, ADMX-HF \cite{Stern:2016bbw} or HAYSTAC \cite{Zhong:2018rsr}
 which explicitly depend on the axion density. However, in other type of experiments such as CASPEr \cite{JacksonKimball:2017elr} or QUAX \cite{Barbieri:2016vwg}
   when the observables are proportional to the axion velocity 
  $\mathbf{v_a}\sim\boldsymbol{\nabla} a$ the AQN-induced relativistic  axions will clearly enhance the signal,  as shown by Eq. (\ref{H}) in Appendix \ref{broadband}. 
  In that case, it is the axion flux $\Phi^{\rm AQN}_a$, proportional to axion velocity
  $\mathbf{v_a}$, rather than the axion density becomes 
  the relevant observable. In any case, axion density and flux are closely related: $ v_a \rho_a^{\rm AQN}\approx m_a\Phi^{\rm AQN}_a$, and we present our basic numerical  results in the main body of the text in terms of both.
  
We would like to emphasize that the main result of this work is the computation of the annual, daily modulations of the induced axions,  as well as rare  bursts-like  amplifications. While the annual  modulations have been discussed in the literature   \cite{Freese:1987wu,Freese:2012xd}, the daily modulations are normally ignored in DM literature because in WIMPs based models the effect is negligible. This is no longer the case with the AQN model when both effects are large, of the order of $10\%$.
More specifically, we computed the AQN-induced axion flux on the Earth's surface, which can be conveniently presented as follows
\be
\label{flux}
\la E_a\ra \Phi^{\rm AQN}_a(t)\simeq 10^{14}A(t) \left[{\rm\frac{eV}{cm^2s}}\right], ~~~ \la E_a\ra\simeq 1.3\,m_a, ~~
\ee
where $A(t)$ is the modulation/amplification time dependent  
factor. We normalize the flux such that $\la A(t)\ra =1$ if averaged over very long period of time, much longer  than few years.

The effect of the annual modulation has been known since   \cite{Freese:1987wu,Freese:2012xd}.
For the AQN-model we computed  the annual modulation  parameter $\kappa_{\rm (a)}$     
defined as follows:
  \begin{equation}
\label{eq:annual}
 A_{\rm (a)}(t)\equiv[1+\kappa_{\rm (a)} \cos\Omega_a (t-t_0)],
\end{equation}
 where $\Omega_a=2\pi\,\rm yr^{-1} $  is the angular frequency of the annual modulation
 and   label $``a"$ in $\Omega_a $ stands for annual.   The $\Omega_a t_0$ is the phase shift corresponding to the maximum on June 1 and minimum on December 1 for the standard galactic DM distribution.   
 
We have also computed  the daily modulations with parameters defined as follows: 
 \begin{equation}
\label{eq:daily}
A_{\rm (d)}(t)\equiv[1+\kappa_{\rm (d)} \cos(\Omega_d t-\phi_0)],
\end{equation}
 where   $\Omega_d=2\pi\,\rm day^{-1}$ is the angular frequency of the daily modulation, while $\phi_0$ is the phase shift similar to $\Omega_at_0$ in (\ref{eq:annual}). It   can be assumed to be constant on the scale of days. However, it actually slowly changes  with time due to the  variation of the direction of  DM wind   with respect  to the Earth. In addition to annual and daily modulations, we also show that the factor $A(t)$
can be numerically large  for rare  bursts-like events,  
the so-called ``local flashes''.   These  short  
bursts   resulting from the interaction of the AQN hitting  the Earth in  a close vicinity of a detector.

It is important to emphasize  that, in  the present work, along with the accompanying paper \cite{Budker:2019zka}, we are studying   the relativistic axions, $v_a\sim0.6c$, which represent a  \emph{direct} manifestation of the AQN model. It should be contrasted with 
  \emph{indirect} manifestations of the AQN model mentioned above. 
An observation of axions with very distinct spectral properties compared to those predicted from conventional galactic axions  with $v_a\sim 10^{-3}c$ would be a smoking gun for the AQN framework and this may then answer a fundamental question on the nature of DM.  

The presentation is organized as follows. 
We start   with a brief overview of the axion emission mechanism in the AQN framework in Sec. \ref{sec:the AQN model and its axion emission mechanism}.   In Sec. \ref{sec:potential enhancements} we compute the relevant parameters describing the modulations and amplifications as announced in this Introduction. The corresponding derivation of analytical equations and algorithm of numerical simulation are left to Sec. \ref{sec:annihilation modeling} and \ref{sec:algorithm and simulation}, with detailed results presented in \ref{sec:discussion of results}. Finally, we conclude with some thoughts on possible future developments in Sec. \ref{sec:conclusions and future directions}.

\section{The AQN model and its axion emission mechanism}
\label{sec:the AQN model and its axion emission mechanism}
The AQNs hitting the Earth surface is given by \cite{Lawson:2019cvy}:
\be
\label{eq:D Nflux 3}
%\begin{aligned}
\frac{\langle\dot{N}\rangle}{4\pi R_\oplus^2}
=\frac{0.4}{{\rm  {km^{2}}yr }}\left(\frac{10^{24}}{\langle B\rangle}\right)
\left(\frac{\rho_{\rm DM}}{0.3{\rm \frac{GeV}{cm^3}}}\right)
\left(\frac{\la v_{\rm AQN} \ra}{\rm 220\kmps}\right).~~~ 
%\end{aligned}
\ee
Eq. \eqref{eq:D Nflux 3} shows that conventional DM detectors are too small to detect AQNs directly. 
However, axions will be emitted when the AQN crosses the  Earth interior, due to the annihilation processes that will lead to time-dependent perturbations of the axion DW.
 The time-dependent perturbations due to annihilation processes will change the equilibrium configuration of the axion DW shell, and axions will be emitted because the total energy of the system is no longer at its minimum when some portion of the baryon charge in the core is annihilated. To retrieve the ground state, an AQN will therefore lower its domain wall contribution to the total energy by radiating axions. The resulting emitted axions can be detected by conventional haloscope  axion search experiments.
 
 The emitted axion velocity spectrum was calculated in \cite{Liang:2018ecs} using the following approach: Consider a general form of a domain wall:
\begin{equation}
\label{eq:2.3 phi soln}
\phi(R_0)=\phi_w(R_0)+\chi
\end{equation}
where $R_0$ is the radius of the AQN, $\phi_w$ is the classical solution of the domain wall, and $\chi$ describes excitation due to the time-dependent perturbation. $\phi_w$ is the DW time-independent classical solution, while $\chi$ describes on-shell propagating axions. Suppose an AQN is travelling in the vacuum where no annihilation events are taking place. The DW solution will remain in its ground (minimum energy) state $\phi(R_0)=\phi_w(R_0)$. Since there is no excitation (i.e. $\chi=0$), no free axion can be produced. When some baryon charge of the AQN is annihilated, the AQN starts loosing mass, its size decreases from $R_0$ to a slightly smaller radius $R_{\rm new}=R_0-\Delta R$. The quantum state $\phi(R_0)=\phi_w(R_0)$ is then no longer the ground state, because a lower energy state $\phi_w(R_{\rm new})$ becomes available. The state of the domain wall then becomes $\phi(R_0)=\phi_w(R_{\rm new})+\phi_w'(R_{\rm new})\Delta R$, and the domain wall acquires a nonzero exciting mode $\chi=\phi_w'(R_{\rm new})\Delta R$ leading to the production of free axions. Thus, whenever the domain wall is excited, corresponding to $\chi\neq0$, freely propagating axions will be produced and emitted by the excited modes.
The emission of axions is therefore an inevitable consequence of the annihilation of antimatter AQNs and of AQNs minimizing their binding energy. We refer the readers to Refs.  \cite{Liang:2018ecs} for the technical details of the emission mechanism and the calculation of its axion spectrum. For convenience, we also review  the important  results from \cite{Liang:2018ecs} in Appendix  \ref{app:spectral properties in the rest frame}.

Let $\rmd N/\rmd B$ be the number of AQNs which carry the baryon charge [$B$, $B+dB$].
We shall use the same models for $\rmd N/\rmd B$ as in \cite{Lawson:2019cvy}, and we refer to that paper for a description of the baryon charge distributions still allowed. Following \cite{Lawson:2019cvy}, the mean value of  the baryon charge $\langle B\rangle $ is given by 

\begin{equation}
\label{eq:f(B)}
\langle B\rangle 
=\int_{B_{\rm min}}^{10^{28}}\rmd B~B f(B),
\qquad f(B)\propto B^{-\alpha}
\end{equation}
where $f(B)$ is properly normalized distribution  and the  $\alpha$  is power-law index which assumes the following values:

\begin{equation}
\label{eq:2.2 f(B) ass_alpha}
\alpha=2.5,~2.0,~{\rm or}\left\{
\begin{aligned}
&1.2  &B\lesssim 3\times 10^{26}  \\
&2.5  &B\gtrsim  3\times 10^{26}\ .
\end{aligned}\right.
\end{equation}

One should   note that that the algebraic scaling (\ref{eq:f(B)}) is a generic feature of the AQN formation mechanism based on percolation theory  \cite{Ge:2019voa}. The parameter $\alpha$ is determined by the properties of the domain wall formation during the QCD transition in the early Universe, but it cannot be theoretically computed in strongly coupled QCD. Instead, the parametrization  (\ref{eq:2.2 f(B) ass_alpha}) is based on  fitting the observations of the Extreme UV emission from the solar corona as discussed  in \cite{Raza:2018gpb,Lawson:2019cvy}.

Another parameter that defines  the distribution function (\ref{eq:f(B)}) is the minimum baryonic charge $B_{\rm min}$. 
We use the same models as discussed in  \cite{Raza:2018gpb,Lawson:2019cvy} and take $B_{\rm min}=10^{23}$ and $B_{\rm min}=3\times10^{24}$. Therefore, we have a total of 6 different models for $f(B)$.    In Table \ref{tab:mean B} we show the mean baryon charge $\langle B \rangle$ for each of the 6 models.
\begin{table} [h] %[h] here; [t] top; [b] bottom
	\caption{Values of the mean baryon charge $\langle B\rangle$ for different parameters of the AQN mass-distribution function.} % title of Table
	\centering % used for centering table
	\begin{tabular}{c|ccc} % centered columns (4 columns)
		\hline\hline
		$(B_{\rm min},\alpha)$ & 2.5                 & 2.0                 & (1.2, 2.5)          \\ \hline
		$10^{23}$                   & $2.99\times10^{23}$ & $1.15\times10^{24}$ & $4.25\times10^{25}$ \\ 
		$3\times10^{24}$                   & $8.84\times10^{24}$ & $2.43\times10^{25}$ & $1.05\times10^{26}$ \\ \hline\hline
	\end{tabular}
	\label{tab:mean B} % is used to refer this table in the text
\end{table}
For simulations in this work, we will only investigate parameters that give $\langle B\rangle\gtrsim10^{25}$ to be consistent 
with IceCube and Antarctic Impulsive Transient Antenna  (ANITA) experiments as discussed in  \cite{Lawson:2019cvy}.
 Therefore, in our numerical studies we  exclude two models corresponding  $B_{\rm min}\sim10^{23}$   with power-law index  $\alpha=2.5$ and $\alpha=2.0$.

\section{modulation and  amplifications: WIMPs and the AQN-induced axions}
\label{sec:potential enhancements}

In this section we quantify the time modulations and amplifications of the axion flux that can potentially be realized in the AQN model, and compare it to the conventional DM candidates
such as weakly interacting massive particles
(WIMPs). We examine time modulations and possible enhancements in descending order of importance, from annual to daily modulation effects \cite{Freese:1987wu,Freese:2012xd} specific to the AQN model as well as other new phenomena.

Our results are presented in Table \ref{tab:potential enhancement}, and the resulting comparison between AQN-induced axions and conventional DM enhancements can be summarized as follow: 

The annual modulation discussed in Sec. \ref{subsec:annual modulation} has a similar amplitude in both cases, but the daily modulation presented in Sec. \ref{subsec:daily modulation} is much stronger for AQN-induced axions than for conventional WIMP-like models. Sec. \ref{subsec:statistical fluctuation} and \ref{subsec:local flashes} introduce two new, time-dependent phenomena, that are unique to the AQN model, and do not exist in conventional DM: the statistical fluctuations and the ``local flashes" respectively.
The latter effect  becomes operational when the axion detector happens to be in the vicinity of the point where AQN enters or exits 
the Earth surface. We consider this novel phenomenon as the most promising effect  which may drastically enhance the discovery potential for the axion search experiments and provide a decisive test of the AQN model. Gravitational lensing is another modulation effect, originally discussed in \cite{Patla:2013vza,Bertolucci:2017vgz}. We revisit this effect in subsection \ref{subsec:gravitational lensing} assuming a conventional galactic DM distribution (in coordinate  and momentum spaces) which explains our result of a negligible effect in comparison with huge enhancement reported in \cite{Patla:2013vza,Bertolucci:2017vgz} where nonconventional streams of DM were considered.

\begin{table}[h]
\captionsetup{justification=raggedright}
	\caption{Comparison of potential enhancement: AQN-induced axions vs. WIMPs. Amplifications factors are listed up to order of magnitude estimate.} % title of Table
	\centering % used for centering table
	\begin{tabular}{ccc}
		\hline\hline
		Potential enhancement  &WIMPs & \begin{tabular}{@{}c@{}}AQN-induced \\ axions\end{tabular} \\\hline
		Annual modulation         & $1\%-10\%$ \cite{Freese:1987wu,Freese:2012xd}   &  $1\%-10\%$         \\
		 Daily modulation           & $\ll1\%$ \cite{Freese:2012xd}   &  $1\%-10\%$          \\
		 Statistical fluctuation     &   0       &  $20\%-60\%$          \\
		 Local flashes                  &0      &   $10^2-10^3$             
		\\\hline
		 Gravitational lensing     & $10^4-10^6$ \cite{Patla:2013vza,Bertolucci:2017vgz} 
		 &  $\lesssim1\%$ \\\hline \hline
	\end{tabular}
	\label{tab:potential enhancement}
\end{table}

\subsection{Annual modulation}
\label{subsec:annual modulation}

Conventional DM candidates such as galactic axions and WIMPs are affected by annual modulation with change estimated to be $\mathcal{O}(1\%-10\%)$ based on standard halo model (SHM) of the galactic halo  \cite{Freese:1987wu,Freese:2012xd}. To understand this result, we first note the local speed of DM stream is not constant and subject to an annual modulation due to the motion of the Earth  \cite{Freese:1987wu,Freese:2012xd}:
\begin{equation}
\label{eq:mu(t)}
\mu(t)\simeq V_\odot+bV_\oplus\cos\omega_{\rm a}(t-t_0)
\end{equation}
where $V_\odot=220\kmps$ is the oribital speed of the Sun around the galactic center, $V_\oplus=29.8\kmps$ is the orbital speed of the Earth around the Sun, $\omega_{\rm a}=2\pi\,\rm yr^{-1}$ is the angular frequency of the annual modulation, and $|b|\leq1$ is a geometrical factor associated with the direction of local velocity $\boldsymbol{\mu}$ relative to the orbital plane of Earth. Hence, it is natural to expect the amplification must be of order $\mathcal{O}(V_\oplus/V_\odot)\sim10\%$, as the incoming flux of particles depends on the incident speed $\mu$. 

Similar arguments apply to the AQN-induced axions. We first note the flux of AQN-induced axions (derived in Sec. \ref{subsec:the axion flux density on Earth's surface}) is:
\begin{equation}
\label{eq:m_a Phi_a simeq}
m_a\Phi_a
\simeq
\frac{v_a}{c}\frac{\langle\dot{N}\rangle\langle\Delta m_{\rm AQN}\rangle}{16\pi R_\oplus^2}\ ,
\end{equation}
where $v_a\sim0.6c$ is speed of the emitted axions, 
$\langle\dot{N}\rangle$ is the expected hit rate of AQNs on Earth, and $\langle\Delta m_{\rm AQN}\rangle$ is the total average mass loss\footnote{For sake of brevity, in this work we adopt the following terminology: whenever ``per AQN'' is referred, we mean ``per $\langle B\rangle$ baryon charge''.} per single AQN  
with baryon charge $B$ such that $\Delta m_{\rm AQN}\approx m_p \Delta B$. Numerically, the corresponding flux
of the AQN-induced axions is 
\begin{equation}
\label{eq:E_a Phi_a numerical}
\langle E_a\rangle\Phi_a
\sim 10^{14}\left[\rm\frac{eV}{cm^2s}\right]\,,
\end{equation}
as the presented in Eq. \eqref{flux} with $\langle A(t)\rangle=1$.

We should mention here  that the exact  formula to be derived in Sec. \ref{subsec:the axion flux density on Earth's surface} contains some additional features (e.g. angular dependence), but it is sufficient to use Eq. \eqref{eq:m_a Phi_a simeq} for qualitative discussion in this section.
The linear relation \eqref{eq:m_a Phi_a simeq} simply states that the 
the output axion flux rate $m_a\Phi_a$ is proportional to the amount of AQN flux supplied $\langle\dot{N}\rangle$ and its mass loss $\langle\Delta m_{\rm AQN}\rangle$. Clearly, the hit rate $\langle \dot{N}\rangle$ is, by definition, linearly proportional to the magnitude of incident speed $\mu$, resulting in a modulation up to order $\mathcal{O}(V_\oplus/V_\odot)\sim10\%$. On the other hand, we note the mass loss $\langle\Delta m_{\rm AQN}\rangle$ is a speed independent quantity by its conventional definition:

\begin{equation}
\label{eq:dm_AQN}
\rmd m_{\rm AQN}
=-\sigma\rho v\rmd t
=-\sigma\rho \rmd s\ ,
\end{equation}
where $\sigma$ is the effective cross section of the AQN, $\rho$ is local density of the Earth, $v$ is the speed of the AQN, $t$ and $s$ is propagation time and the path length of the AQN respectively. Thus, the farther an AQN travels underground the more axions emitted due to mass loss, but this is insensitive to the AQN speed. 

\begin{figure}
	\centering
	\captionsetup{justification=raggedright}
	\includegraphics[width=1\linewidth]{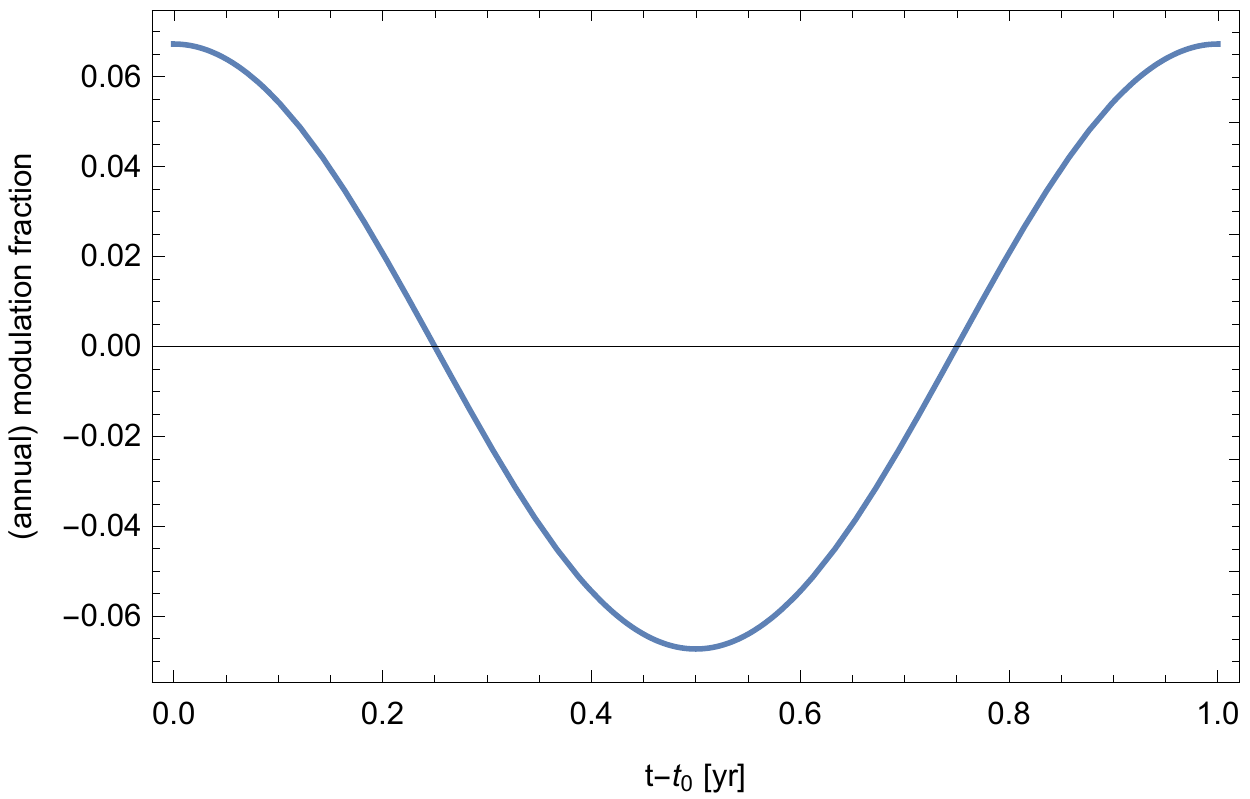}
	\caption{Annual modulation fraction as a function of time $(t-t_0)$, the geometrical factor is chosen to be $b=0.497$ \cite{Freese:2012xd}.}
	\label{fig:annual modualation}
\end{figure}

To illustrate this, we plot the annual modulation fraction, defined as the size of the modulation amplitude relative to its average flux density, as a function of time in Fig. \ref{fig:annual modualation}. In this work, we choose two extreme cases $\mu=V_\odot\pm V_\oplus$ to demonstrate the existence of annual modulation in our numerical simulations.

 \subsection{Daily modulation}
\label{subsec:daily modulation}

The effect of daily modulation is rarely considered important in conventional discussion of DM candidates because the additional correction (at most $0.5\kmps$ rotational velocity on the surface of the Earth) is much smaller than the Earth's orbital velocity $V_\oplus$. However, the scenario is drastically different in case of axion emission from AQNs. The basic reason is that the AQNs are the  composite microscopical objects\footnote{This should be  contrasted with conventional DM candidates which are microscopic fundamental particles characterized by a  cross section that is independent from orientation and position of the Earth with respect to Sun and galactic centre.} which lose a portion of their material while crossing the Earth. As a result the size of an AQN decreases while it crosses the Earth. 

\begin{figure}[h]
	\centering
	\captionsetup{justification=raggedright}
	\includegraphics[width=0.9\linewidth]{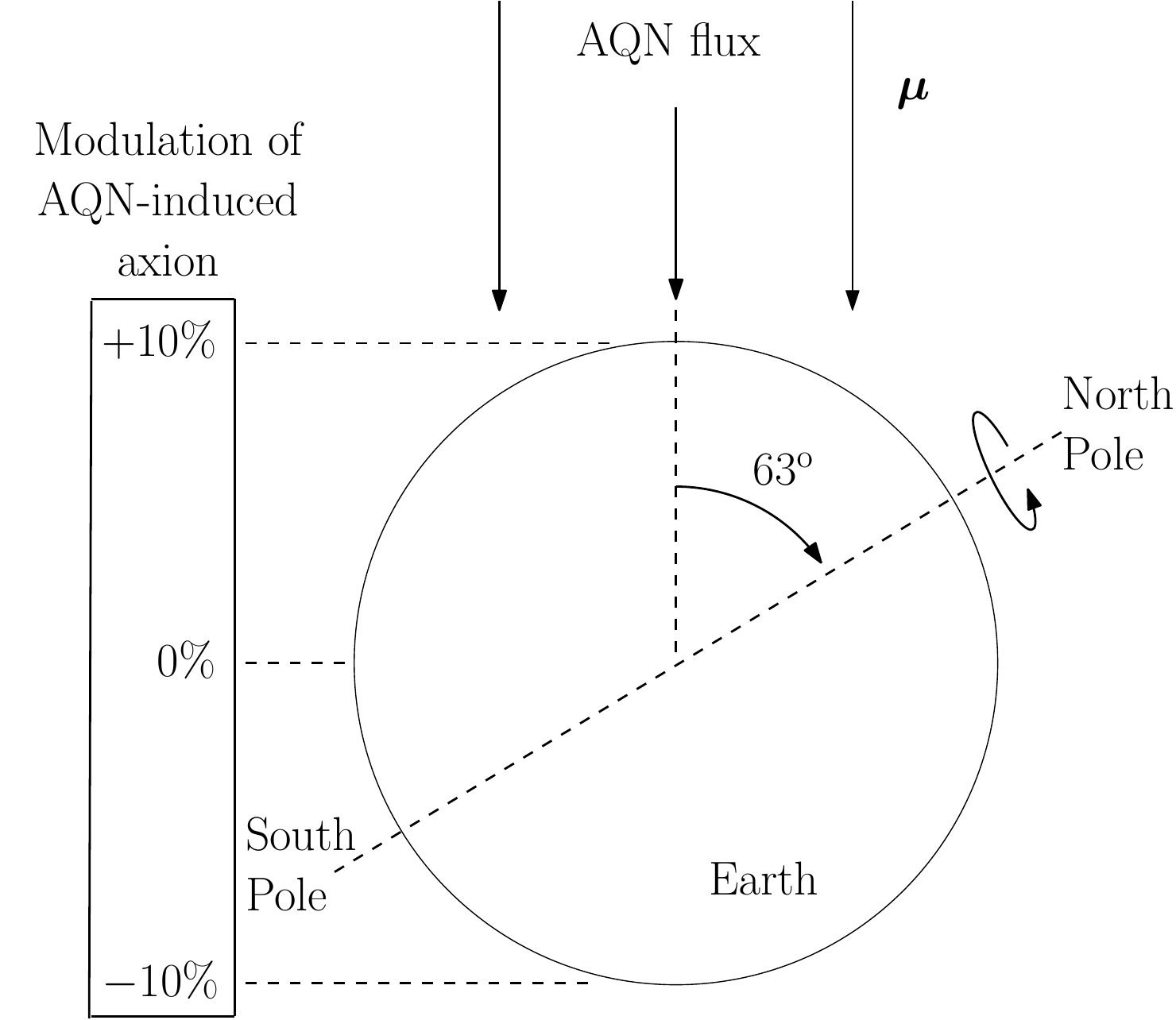}
	\caption{Mechanism of daily modulation for the AQN-induced axions. The DM wind in form of the AQNs  comes at a fixed angle $63\degree$ due to the alignment of the celestial equator relative to the galactic plane. Consequently, about 10\% more (less) axions are emitted in the upper (lower) hemisphere compared to the average value. Due to self-rotation of the Earth, daily modulation of order $10\%$ is expected. The effect is stronger for    detectors built in lower latitude region.}
	\label{fig:daily modulation}
\end{figure}

The basic mechanism of the daily modulation of the AQN-induced axions can be describe as follow: Eq. \eqref{eq:dm_AQN} implies that the flux of AQN-induced axions is sensitive to the local mass loss of AQNs. Furthermore, the same equation suggests that the local mass loss is not constant 
during  the passage of the AQN through Earth as the cross section $\sigma$ decreases  along the trajectory. Consequently, the annihilation rate, and therefore the  axion production rate, also decreases along the trajectory. Since the AQN cross section decreases as they traverse Earth and lose mass, there is then more axions emitted facing the wind compared to facing away from it. Fig. \ref{fig:daily modulation} illustrates this mechanism: the AQN flux impacts the Earth at a fixed angle $63\degree$ due to the alignment of the celestial equator (the plane of Earth's equator) relative to the Galactic plane. Consider  the downward motion as  shown  in Fig. \ref{fig:daily modulation}.   For this case  we expect more heat (and therefore more axions) to be emitted in the upper hemisphere (the half sphere facing the AQN wind) than the lower one (the half sphere opposing the AQN wind). The difference can be as large as $\sim10\%$ as we will soon estimate. Then, as the Earth rotates daily, we expect to see a daily modulation of order $10\%$ especially for detectors built in lower latitude regions with respect to the north pole.

In order to estimate the amplitude of this effect, we consider the ratio between the annihilation cross section when an AQN enters and when AQN exits the Earth (assuming the same orientation for the wind $\boldsymbol{\mu}$):
\begin{equation}
\label{eq:dm_AQN m_AQN}
\frac{\sigma_{\rm entry}}{\sigma_{\rm exit}}\simeq \frac{\sigma_{\rm entry}}{(\sigma_{\rm entry}-\Delta\sigma)}
\simeq \left(1+\frac{\Delta\sigma}{\sigma}\right)
\simeq 1+\frac{2}{3}\frac{\Delta B}{B}\ ,
\end{equation}
where we use the relation $\sigma \propto R^2  \propto B^{2/3}$. 
Therefore, the total fluctuation of daily modulation deviated from the mean value is half of the above estimation
\eqref{eq:dm_AQN m_AQN}:
\begin{equation}
\label{eq:percetage fluctuation}
{\rm daily~ modulations}\equiv \kappa_{\rm (d)}
\simeq\frac{1}{3}\frac{\langle\Delta B\rangle}{\langle B\rangle}\ ,
\end{equation}
which can be trusted as long as the factor on the right hand side is numerically small and   expansion (\ref{eq:dm_AQN m_AQN}) is justified. 
From Monte Carlo simulation, we calculated that the typical fraction of total mass loss $\langle\Delta B\rangle/\langle B\rangle$ is about 30\%. Therefore, we expect an amplitude modulation of order 10\% from the mean value in mass loss. The estimate \eqref{eq:percetage fluctuation} is consistent  with numerical simulations, later demonstrated in Sec. \ref{sec:discussion of results}, which supports our interpretation in terms of the daily modulation. The corresponding daily  modulation can be described with the parameter $A_{\rm (d)}(t)$  as defined in Eq. \eqref{eq:daily}. 
%by definition \eqref{eq:A(t)}:
%\begin{equation}
%\label{eq:daily}
%A_{\rm d}(t)\equiv1+\kappa_{\rm d} \cos(\omega_{\rm d} t-\phi_0),
%\end{equation}
%where   $\omega_{\rm d}=2\pi\,\rm day^{-1}$ is the angular frequency of the daily modulation, while $\phi_0$ is the phase shift similar to $t_0$ in (\ref{eq:mu(t)}). 
$A_{\rm (d)}(t)$ can be assumed to be constant on the scale of days. However, it actually slowly changes  with time due to the  variation of the direction of  DM wind   with respect  to the Earth's position and orientation.

Lastly, we would like to point out another interesting consequence: there also exists a ``spatial" modulation: axion flux is slightly more intense (the same $10\%$) in the Northern Hemisphere compared to the south, because the DM wind points to the northern portion of  the Earth, see Fig. \ref{fig:daily modulation}.

\subsection{Statistical fluctuation}
\label{subsec:statistical fluctuation}
In this subsection we study a new type of time-dependent  DM signal which has not been discussed before in the context of conventional DM: AQNs are macroscopically large objects with a number density approximately 23 orders of magnitude  smaller in comparison to WIMPs with mass $\sim 10^2$ GeV. Therefore Poisson fluctuations, which are completely irrelevant for conventional DM, could potentially be important for the AQN model. The tiny  flux (\ref{eq:D Nflux 3}) of AQNs is an explicit manifestation of this unique feature. The details of the AQN statistical fluctuation depend on the AQN size (and mass) distribution, trajectories and velocities.

In order to study the statistical fluctuation of AQN-induced  axions, we consider the flux formula \eqref{eq:m_a Phi_a simeq} and conjecture that the two quantities, $\langle\dot{N}\rangle$ and $\langle\Delta m_{\rm AQN}\rangle$, may have large statistical fluctuation due to the low number statistics. The average hit rate of AQNs hitting the Earth is about $1\rm\,s^{-1}$ \cite{Lawson:2019cvy}, while the crossing time of an AQN inside the Earth interior  is of order $\Delta t\sim R_\oplus/v_{\rm DM}\sim30\rm\,s$. Hence, we estimate that there are only about 30 AQNs in the Earth interior at any given moment. Statistical fluctuation can therefore be  important for AQNs, unlike conventional DM.

We numerically simulate the statistical fluctuations using a two-step Monte Carlo. First, we simulate $N$, the number of AQNs in the Earth interior    at a given moment, by Poisson distribution:
\begin{equation}
\label{eq:Prob(N)}
\begin{aligned}
{\rm Prob}(N)
\sim
\frac{\lambda^N}{N!}e^{-\lambda},
\quad\lambda=\langle\dot{N}\rangle\langle\Delta t\rangle\ ,
\end{aligned}
\end{equation}
where $\langle\dot{N}\rangle=0.672{\rm~s}^{-1}$ is the average hit rate of AQNs \cite{Lawson:2019cvy}, and $\langle \Delta t\rangle\sim30$ s is the average time duration of an AQN inside the Earth\footnote{A subtle point is, $\langle\dot{N}\rangle$ is also proportional to $\mu$, the mean speed of AQN. Therefore for some model parameters (e.g. annual modulation) that modify $\mu$, $\langle\dot{N}\rangle$ should be also modified correspondingly.}. And then, we obtain $\Delta m_{\rm AQN}$, the average mass loss per AQN from this sample size $N$. By repeating the above process, we obtain the standard deviation of $\langle\Delta m_{\rm AQN}\rangle$ where fluctuation in $\langle\dot{N}\rangle$ is already accounted.

The results are  shown in the last columns of Table \ref{tab:summary of statistical fluctuations}. We find statistical fluctuations of the order the ${\cal O}(20\%-60\%)$, depending on the size distribution model  (\ref{eq:f(B)}), (\ref{eq:2.2 f(B) ass_alpha}).  Numerically, the effect is greater than the enhancements assessed in the previous subsections for annual and daily modulations. 

It is interesting to  observe that smaller values of power-law index $\alpha$ corresponds to a larger fluctuation of mass loss $\langle\Delta m_{\rm AQN}\rangle$. Qualitatively this can be explained by the fact that a smaller value of $\alpha$ 
corresponds to higher AQNS average mass and consequently the rate of AQNs hitting Earth is reduced. This implies a larger dispersion, in agreement with explicit numerical computations presented in Table \ref{tab:summary of statistical fluctuations}. In addition, we note that the statistical fluctuations of $\langle\Delta m_{\rm AQN}\rangle$ are almost insensitive to the AQN mean speed $\mu$ (DM wind).  

\begin{table*} [!htp]%[h] here; [t] top; [b] bottom
\captionsetup{justification=raggedright}
	\caption{Summary of statistical fluctuations ($B_{\rm min}=3\times10^{24}$ unless specified, $10^3$ trials involved in the two-step Monte Carlo). The uncertainties represent $1\sigma$ significance. $\sigma_m$ is the standard deviation of $\langle \Delta m_{\rm AQN}\rangle$ in the simulation.} % title of Table
	\centering % used for centering table
	\begin{tabular}{ccccccc} % centered columns (4 columns)
		\hline\hline %inserts double horizontal lines
		$\langle B\rangle$ &$\alpha$ &Other parameters & $\langle\Delta t\rangle$ [s] & $\langle\dot{N}\rangle\langle\Delta t\rangle$ & $\langle \Delta m_{\rm AQN}\rangle$ [kg] & $\sigma_m/\langle \Delta m_{\rm AQN}\rangle$  \\\hline
		$8.84\times10^{24}$ &2.5  &--      & $38.3\pm5.6$                 &   $25.8\pm5.0$                                     &  $(4.93\pm1.26)\times10^{-3}$    &25.6\%                             \\
		$8.84\times10^{24}$ &2.5  &$\mu=V_\odot - V_\oplus$      & $41.8\pm8.5$                 &  $25.2\pm5.0$                                      &  $(4.90\pm1.17)\times10^{-3}$    &  23.8\%                                   \\
		$8.84\times10^{24}$ &2.5  &$\mu=V_\odot + V_\oplus$      &  $35.2\pm4.5$                &  $26.1\pm5.1$                                       &  $(4.96\pm1.34)\times10^{-3}$    &  27.1\%                                  \\
%		$8.84\times10^{24}$ &2.5  &$\epsilon=3$      & $105.1\pm9.0$                &  $70.6\pm8.3$                                       &  $(1.34\pm0.52)\times10^{-2}$   &  $38.8\%$                                   \\
		$8.84\times10^{24}$ &2.5  &Solar gravitation      & $32.5\pm3.8$                &  $25.0\pm4.9$                                       &  $(4.97\pm1.61)\times10^{-3}$   &  32.5\%                                   \\
		$2.43\times10^{25}$ &2.0  &--     & $37.5\pm4.8$                     & $25.2\pm5.0$                                         & $(8.13\pm4.51)\times10^{-3}$  &  55.5\%             \\
		$4.25\times10^{25}$ &(1.2, 2.5) & $B_{\rm min}=10^{23}$       & $62.3\pm15.2$                & $41.9\pm6.6$                                     & $(1.03\pm0.49)\times10^{-2}$          &47.6\%                             \\
		$1.05\times10^{26}$ &(1.2, 2.5) &--     & $35.4\pm4.7$                     & $23.8\pm4.9$                                         & $(2.48\pm0.98)\times10^{-2}$              &39.5\%                           \\\hline\hline
	\end{tabular}
	\label{tab:summary of statistical fluctuations} % is used to refer this table in the text
\end{table*}

\subsection{Local flashes}
\label{subsec:local flashes}
Now we turn to the most interesting and most promising enhancement effect. Sharing a similar origin to the statistical fluctuation, we note that the detection signal of axion flux emitted by AQNs may be greatly amplified via a ``local flash'' on rare occasions when an AQN hits (or exits) the Earth surface in the vicinity of an axion search detector. 

To understand the local flash, we first note that the intensity of axion flux is inversely proportional to distance square from the source. As estimated from the preceding section, there are only about 30 AQNs inside the Earth at any moment. Most of these AQNs do not loose much momentum and they travel a distance  of order $R_\oplus$ along straight  path without changing their original  directions. Now, consider a case when an  AQN is moving from a distance $d$ close enough to the axion detector. The intensity of the axion flux will be greatly enhanced by factor of $(R_\oplus/d)^2$ for  a short period of time. This short enhancement in intensity is called a ``local flash".  In the following, we will estimate the signal amplification factor of a local flash compared to the average axion flux induced by AQNs, and we will derive the instrumental requirements for a possible detection.

\begin{figure}[h]
	\centering
	\captionsetup{justification=raggedright}
	\includegraphics[width=0.8\linewidth]{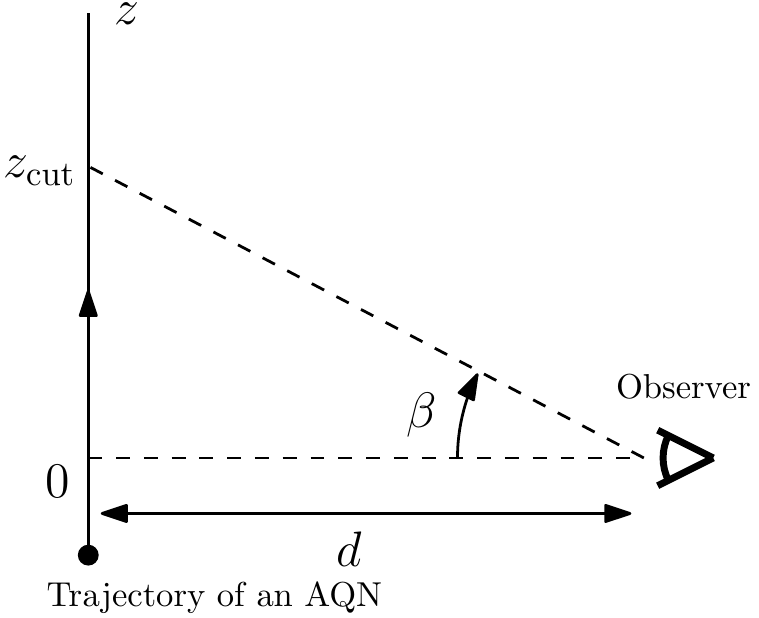}
	\caption{Cause of local flashes. Since the mean free path of AQNs in Earth is of order $R_\oplus$, the axion flux is amplified by $R_\oplus^2/d^2$ in short time (within $z_{\rm cut}\sim d$) when an AQN moves close to the detector by distance $d$.}
	\label{fig:coord local flashes}
\end{figure}

As shown in Fig. \ref{fig:coord local flashes}, suppose an AQN is moving close to the detector with a minimum distance $d$. The intensity of the emitted axion flux is maximized within $z\lesssim d$. Assuming the time is sufficiently short, the mass loss rate $\dot{m}_{\rm AQN}$ and velocity $\dot{z}\simeq v_{\rm AQN}$ can be treated as constants. We obtain the total number (per surface area) of emitted axions $\Delta N_a$ within $z\in[-z_{\rm cut},z_{\rm cut}]$:
\begin{equation}
\label{eq:d Delta N_a dA}
\begin{aligned}
\frac{\rmd}{\rmd S}\Delta N_a
&=\frac{1}{4\pi}\int_{-z_{\rm cut}}^{z_{\rm cut}}
\frac{1}{z^2+d^2}
\frac{\rmd m_{\rm AQN}(z)}{4m_a}  \\
%=\frac{1}{2\pi}\frac{1}{v_{\rm AQN}}\frac{\dot{m}_{\rm AQN}}{4m_a}
%\int_0^{z_{\rm cut}}\frac{\rmd z}{z^2+d^2}  \\
&\simeq\frac{\beta}{2\pi d}\frac{1}{v_{\rm AQN}}\frac{\dot{m}_{\rm AQN}}{4m_a}
\end{aligned}
\end{equation}
where $\beta$ is the angle related to $z_{\rm cut}$ as shown in Fig. \ref{fig:coord local flashes}, and we have implicitly used the relation 
\begin{equation}
\label{eq:dN_a}
\rmd N_a\simeq \frac{1}{4m_a}\rmd m_{\rm AQN}
=\frac{1}{4m_a}\frac{\dot{m}_{\rm AQN}}{v_{\rm AQN}}\rmd z\ ,
\end{equation}
when the integration $\rmd z$ is replaced by integration $\rmd m_{\rm AQN}$, see 
Sec. \ref{subsec:the axion flux density on Earth's surface} for details. Finally, the axion flux density emitted by this AQN is
\begin{equation}
\label{eq:Delta Phi_a(d)}
\begin{aligned}
\Delta\Phi_a(d)
\simeq \frac{v_a}{\tau c}
\frac{\rmd}{\rmd S}\Delta N_a
= \beta\frac{v_a}{m_ac}\frac{\dot{m}_{\rm AQN}}{16\pi d^2}\ ,
%\simeq \frac{v_a}{4m_a}\frac{\beta}{4\pi d^2}
%\frac{\langle \Delta m_{\rm AQN}\rangle}{\langle\Delta t\rangle}
\end{aligned}
\end{equation}
where $\tau\simeq 2d/v_{\rm AQN}$ is the approximate travel time for an AQN inside  the interval $[-z_{\rm cut},z_{\rm cut}]$. 

The expression  \eqref{eq:Delta Phi_a(d)} represents the axion flux emitted by a single AQN travelling at distance $d\ll R_{\oplus}$ from detector.  We want to  compare this ``local flash" with the average flux \eqref{eq:m_a Phi_a simeq} by introducing the amplification factor $A$  defined as the ratio:
\begin{equation}
\label{eq:A(d)}
A(d)\equiv\frac{\Delta\Phi_a(d)}{\Phi_a}
\simeq\frac{\beta}{\langle \dot{N}\rangle\langle \Delta t\rangle}
\left(\frac{R_\oplus}{d}\right)^2\ ,
\end{equation}
where we   approximated $\dot{m}_{\rm AQN}\simeq\langle \Delta m_{\rm AQN}\rangle/\langle\Delta t\rangle$. Using $\langle \dot{N}\rangle\langle\Delta t\rangle\sim30$ (see   Table \ref{tab:summary of statistical fluctuations}), and for $z_{\rm cut}\sim d$, corresponding to  $\beta\sim1$, the typical  amplification becomes significant if  $d\lesssim 0.1 R_\oplus$.  
The time duration $\tau$ of a local flash as a function of amplification $A$ is given by:
\begin{equation}
\label{eq:tau}
\tau
\simeq\frac{2d}{v_{\rm AQN}} 
\simeq\left(\frac{\langle\Delta t\rangle}{\langle\dot{N}\rangle}\right)^{1/2}
A^{-1/2}\ ,
\end{equation}
where we assumed $v_{\rm AQN}\simeq2R_\oplus/\langle\Delta t\rangle$ and $\beta\sim1$ for simplicity. Table \ref{tab:local flashes estimation} shows various values for the time duration $\tau$ as a function of amplification factor $A$. We see that a standard detection signal without amplification ($A\sim1$) lasts for 10 seconds, while a strongly enhanced  signal amplified by $A\sim10^4$ flashes for  0.1 second. However, it is a very rare event as it happens only once in every 5 years. More realistic amplification factors are somewhere between these two limiting cases as    presented in Table  \ref{tab:local flashes estimation}. 

\exclude{
Thus, our conclusion is that any conventional axion search  instrument may benefit from the amplification of local flashes if it  is able to cover the relevant bandwidth for the typical flash duration. For instance, consider an axion with mass $m_a\simeq 10^{-4}{\rm eV}$ corresponding to the frequency $f=\frac{m_a}{2\pi}\simeq 20 ~{\rm GHz}$. The cavity factor is normally $Q_c\simeq 10^5$, while the axion factor $Q_a\simeq 10^6$
such that $Q_a^{-1}$ corresponds to the axion coherence time while $Q_c^{-1}$ counts the cavity storage time.
The bandwidth for conventional cavities is of order $\Delta f\simeq Q_a^{-1} f\sim 20~{\rm  kHz}$. Therefore, the  typical axion   coherence time is  $   (\Delta f)^{-1}\sim 10^{-4}{\rm s}$. 
 These estimates suggest that a typical cavity haloscope is capable of detecting such local flash which lasts $0.3$s with amplification $A\sim 10^3$ according to Table \ref{tab:local flashes estimation}. A required  time-length   of collecting the signal can be achieved with the adequate optimization of the  instrument.   
%  and is able to collect and process the signals within sub-second level.  
}

Let us now estimate the event rate of a local flash for a given amplification $A$.
%, to assure observing a local flash is practical. Fortunately we find it promising to detect a relatively large amplification $A\sim10^2$, and $A\sim10^3$ is still possible for observation on annual basis. 
The probability of observing an AQN for $z\leq d$ is given by:
\begin{equation}
\label{eq:Prob(z_min=d)}
{\rm Prob}(z \leq d)
\simeq\left(\frac{d}{R_\oplus}\right)^2
\simeq\frac{A^{-1}}{\langle \dot{N}\rangle\langle \Delta t\rangle}, \ .
\end{equation}
where we use Eq. (\ref{eq:A(d)}) to express $d$ in terms of $A$. 
The event rate can be expressed in terms of amplification parameter  $A$,
\begin{equation}
\label{eq:Event rate}
\begin{aligned}
{\rm Event~rate}
&=\dot{N}\frac{\tau}{\Delta t}
\cdot{\rm Prob}(z \leq d)  \\
&\simeq\frac{A^{-3/2}}{\sqrt{\langle\dot{N}\rangle\langle\Delta t\rangle^3}},\ .
\end{aligned}
\end{equation}
where averages  $\langle\dot{N}\rangle$ and $\Delta t$ have been numerically computed for different size distribution models, and can be found in Table \ref{tab:summary of statistical fluctuations}. 

Table \ref{tab:local flashes estimation} shows the event rate calculated for a few values of the amplification factor $A$. Specifically, we conclude there is about one event every two days for local flash amplified by $\sim10^2$ if the detector has a time resolution of 1 second. 

In conclusion, the  amplification by local flashes is a unique feature of the AQN-induced  axions. Moreover, these relativistic axions, with $v_a\sim 0.6c$,  have very different spectral properties in comparison to the conventional DM candidates when $v_a\sim 10^{-3}c$.  Therefore, even with an  AQN-induced density (\ref{eq:rho_a intro}) smaller than galactic DM axions the amplification could be sufficiently large because  the AQN-induced flux  could produce   stronger signal  when observables of the experiments is proportional to the axion velocity, see Eq. \eqref{H} in Appendix \ref{broadband} and corresponding discussion in the Introduction. In addition, studying the time correlation of the local flashes can effectively distinguish the true signals from background noise as they are uncorrelated, while such approach does not apply to the conventional axion experiment and most cold DM searches because the distribution of cold DM halo is always uncorrelated and featureless in time. We refer the detailed study of correlation and broadband detection to Ref. \cite{Budker:2019zka}.

\begin{table} %[h] here; [t] top; [b] bottom
\captionsetup{justification=raggedright}
	\caption{Estimation of local flashes: the  time duration, and the corresponding event rate as a function of amplification factor $A$. Here we choose $\beta=1$, $\langle\dot{N}\rangle=0.672\rm\,s^{-1}$ and $\langle\Delta t\rangle=40\rm\, s$.} % title of Table
	\centering % used for centering table
	\begin{tabular}{ccc}
		\hline \hline
		$A$ &  Time Span & Event rate \\ 
		\hline 
		1 & 10 s & 0.3 $\rm min^{-1}$ \\ 
		$10$ & 3 s & 0.5 $\rm hr^{-1}$ \\ 
		$10^2$ & 1 s & 0.4 $\rm day^{-1}$ \\ 
		$10^3$ & 0.3 s & 5 $\rm yr^{-1}$ \\  
		$10^4$ & 0.1 s & 0.2 $\rm yr^{-1}$ \\  
		\hline 
	\end{tabular}
	\label{tab:local flashes estimation}
\end{table}

\subsection{Gravitational lensing}
\label{subsec:gravitational lensing}
This subsection is partly motivated by   Refs. \cite{Patla:2013vza,Bertolucci:2017vgz} where it has been claimed that the Sun or Jupiter can focus the flux of DM particles with speed $\sim(10^{-3}-10^{-1})c$ by an amplification factor up to $\sim10^6$. The claim was  based on two key assumptions:
\begin{enumerate}
	\item The deflection angle $\gamma$ due to gravitational focusing is small, namely $\gamma\ll1$;
	\item The DM flux is colinear.
	%\item The DM particles are non-interacting and can pass through opaque objects such as the Sun and planets.
\end{enumerate}
Assumption 1 requires that the bending angle caused by stars and planets is always very small. Assumption 2 strongly enforces the  gravitational focusing as a result of (assumed) high level of coherency of the DM flux when all particles  move in a highly colinear way with the  same direction. To strengthen the focusing effect, Ref. \cite{Patla:2013vza} additionally assumed the DM particles are non-interacting and can pass through opaque objects such as the Sun and planets, because a transparent Sun has a shorter focal length by one order of magnitude compared to the case of opaque Sun (i.e. for interacting particles) and this results in a correspondingly stronger gravitational focusing correspondingly.
Under these assumptions, calculations are greatly simplified and can be carried out analytically.

It is important to point out that the two assumptions are not well justified in SHM due to the misalignment of the ecliptic plane in the Milky Way. Therefore to satisfy the requirements of perfect focusing alignment and high coherency in propagating direction, Refs. \cite{Patla:2013vza,Bertolucci:2017vgz} considered special streams of slow-moving particles originating from distant point-like sources, such as stars, distant galaxies, and cluster of galaxies, etc.. 

One can imagine that these two assumptions also apply to the case of AQN-induced axions as conjectured in the work \cite{Fischer:2018niu}: Assuming the existence of special emitting source of AQNs for ideal gravitational focusing, one could suspect that the stream of AQNs gains enhancement (up to $\sim10^6$) by gravitational lensing\footnote{In Ref. \cite{Fischer:2018niu}, it is discussed the amplification factor can even go up to $\sim10^{11}$ given the Sun as gravitational lens for source at a very specific distance.}, and consequently the flux of emitted axions is enhanced by the same amplification factor when the AQN stream impacts the Earth. However, the existence of such special emitting source is not of interest in the present work because a SHM is already assumed and implemented. Further investigations carried out in this work shows that in SHM the two assumptions are invalid for AQNs, and therefore the AQN-induced axions. Similarly, the two assumptions are also strongly violated for conventional WIMPs. Therefore, our conclusion is that amplification by gravitational focusing generally do not apply to DM particles in SHM, but the ideas advocated in Refs. \cite{Patla:2013vza,Bertolucci:2017vgz} are not excluded because nonconventional streams of DM were considered. 

\begin{figure}[h]
	\centering
	\captionsetup{justification=raggedright}
	\includegraphics[width=0.8\linewidth]{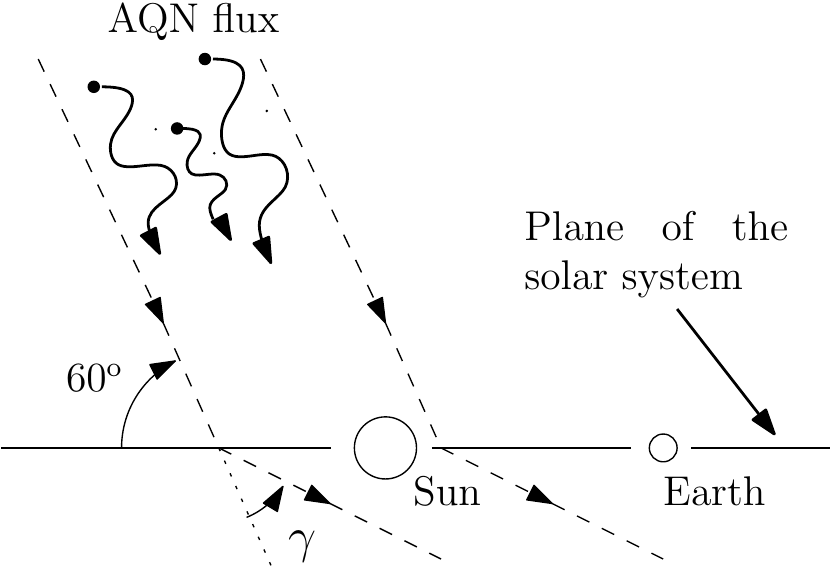}
	\caption{Realistic gravitational deflection of AQN flux. AQN flux is deflected by an angle $\gamma$ due to the gravity of the Sun (or its planet). Since the angle between the ecliptic plane and the galactic equator is always fixed at $60\degree$. The deflection angle $\gamma$ has to be large $\sim60\degree$ in order to make gravitational lensing possible.}
	\label{fig:assumption1}
\end{figure}

In what follows, we clarify the reasons that the two assumptions are violated for SHM. First, assumption 1 is invalid because of the inclination of the ecliptic plane relative to the DM wind: there is a $60\degree$ angle between the ecliptic plane and the Galactic plane, facing the DM wind direction. Consequently, for DM particles to be gravitationally focused to Earth, the deflection angle will have to be of the same order of magnitude. This is illustrated on Fig. \ref{fig:assumption1}. Such large deflection angle are impossible to obtain with lenses like normal stars and planets for the large majority of incoming DM particles, consequently, calculation in Refs. \cite{Patla:2013vza,Bertolucci:2017vgz} is no longer applied as the gravitational lensing is strong. Next, assumption 2 is violated because AQNs (and in fact most DM candidates) have a large velocity dispersion $\sim110\kmps$ that is comparable to their mean galactic velocity. This large velocity dispersion is requirement of the Virial theorem. Thus, the enhancement of DM flux by gravitational lensing as advocated in Refs. \cite{Patla:2013vza,Bertolucci:2017vgz} has a very narrow window for applicability, even for conventional DM candidates like WIMPs, for which both assumptions becomes invalid in the present framework of SHM.

\begin{figure}[h]
	\centering
	\captionsetup{justification=raggedright}
	\includegraphics[width=\linewidth]{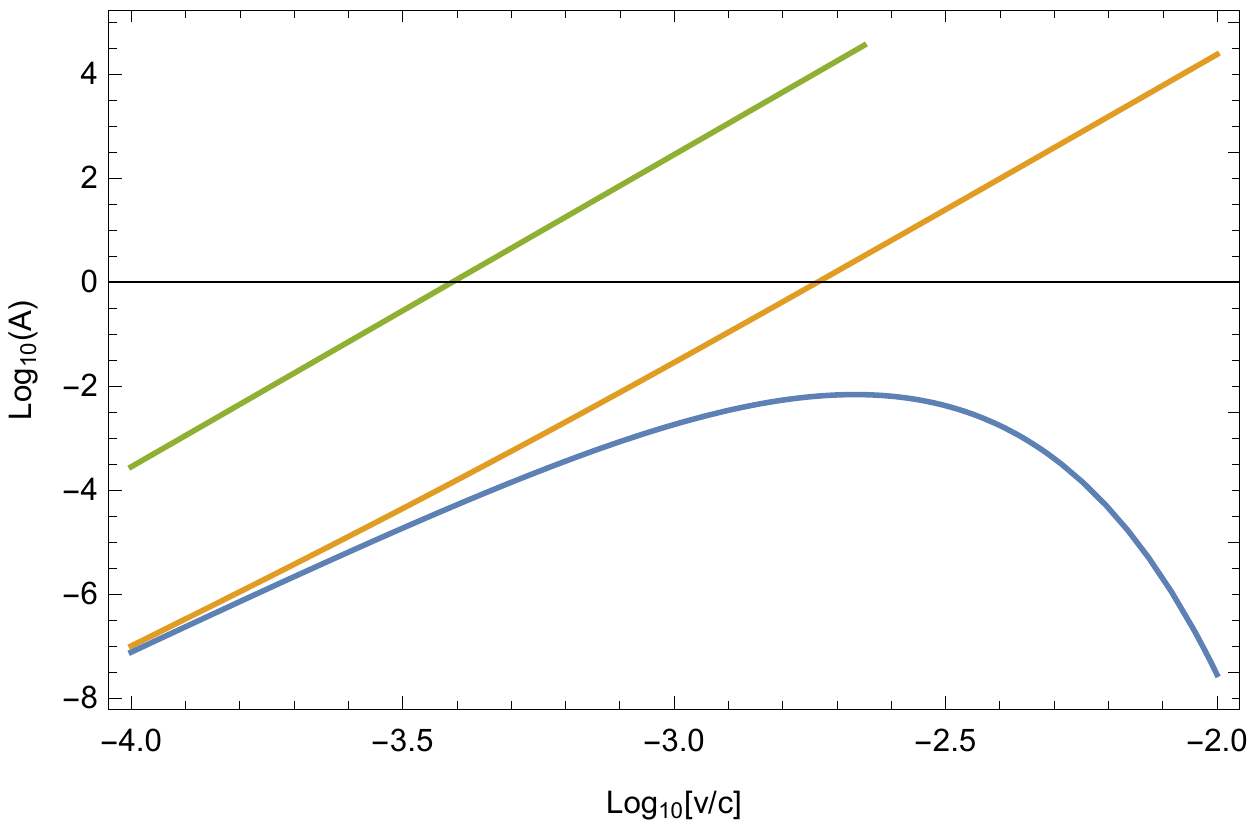}
	\caption{Amplification factor $A(v)$ as a function of AQN velocity spectra. The amplification factor has implicit dependence on $\theta$ and $\sigma$: the angle between its incident direction and the plane of the solar system, and the dispersion in velocity respectively. The mean galactic speed is chosen to be $220\kmps$ with different combination of $(\theta,\sigma)$: (1) $(60\degree,110\kmps)$ in blue, (2) $(0\degree,110\kmps)$ in orange, and (3) $(0\degree,1\kmps)$ in green. The amplification gradually reaches a consistent value estimated in Refs. \cite{Patla:2013vza,Bertolucci:2017vgz} in more idealized conditions.}
	\label{fig:A_max}
\end{figure}

The above arguments have been verified by analytical and numerical calculations, and it is discussed in Appendix \ref{app:detailed calculation of strong gravitational lensing}. Fig. \ref{fig:A_max} shows the resulting enhancement factor $A$ caused by gravitational lensing, where $A$ is plotted as a function of the velocity spectrum of AQN flux. Noting that $A$ implicitly depends on $\theta$ and $\sigma$: the angle between its incident direction and the plane of the Solar system, and the dispersion in velocity respectively.

In the realistic case $(\theta=60\degree,\sigma=110\kmps)$ such that both assumptions are violated, the amplification factor never exceeds $10^{-2}$, so the actual enhancement by gravitational lensing is small and can be neglected in the current study. In addition, we observe the existence of a resonance in the parameter space, namely when $\log_{10}(v/c)\sim -2.7$, the amplification is maximized. This is because the deflection angle is very large and only velocity within a specific range will be deflected within such narrow window $\gamma\simeq60\degree$.

Note that if we considered an   ideal  case  when  the assumption 1 is externally enforced $(\theta=0\degree,\sigma=110\kmps)$ we indeed see some amplification. This happens because the DM wind-solar system alignment   is perfect and the amplification indeed becomes  much stronger  (and grows as a  power law of $v$), see orange line on Fig. \ref{fig:A_max}. Furthermore, if assumption 2 is also enforced
    $(\theta=0\degree,\sigma=1\kmps)$ which is precisely the case considered in  \cite{Patla:2013vza,Bertolucci:2017vgz}, the  AQN flux is perfectly  aligned and highly colinear, see green line on Fig. \ref{fig:A_max}. In this case we indeed    find a strong amplification $\sim 10^4$, in agreement with   computations of  \cite{Patla:2013vza,Bertolucci:2017vgz}. However, it is very hard to imagine how such conditions can be justified, and how a colinear stream of DM could be formed in realistic (not ideal) galactic environment at least in SHM.
    
    The results of the present subsection \ref{subsec:gravitational lensing} were discussed in \cite{Bertolucci:2019jsd}, referring to the amplification $\sim 10^{11}$ mentioned in \cite{Fischer:2018niu}. This factor is well above the enhancement $10^4$ derived in the present work. There is no contradiction: \cite{Fischer:2018niu} did not derive the factor $10^{11}$, but it was mentioned  as a possible strong enhancement motivated by the  analogy with monochromatic coherent electromagnetic emission with $\lambda\simeq 1\,\mu\rm m$   from a distance source.
    This case of a monochromatic EM radiation emitted by a distance source has drastically different physics from the smooth distribution of ordinary DM assumed in present work. 
    Furthermore, the authors of \cite{Bertolucci:2017vgz,Bertolucci:2019jsd} specifically distinguish  ``invisible DM" from ``ordinary DM", the latter being the subject of the present work. This difference is reflected in the assumptions 1 and 2 formulated above. 
    
    To conclude this subsection: we do not expect any substantial amplifications due to the gravitational lensing, at least within  SHM. It is quite possible that future studies accounting for a number of other effects which are ignored in the present SHM (such as accounting for planets) may modify our conclusion. However, this topic is well beyond of the present studies.

\section{Annihilation modeling}
\label{sec:annihilation modeling}
In this section we present all the equations that will be used in our simulations and calculations of the axion-density distribution around the Earth and the flux spectrum of axions passing through  the interior of the Earth.
%\Liang{ Xunyu: ok.{\bf Eric: I suggest to keep as it is because we cited our paper in a proper place.} I realized I claimed something wrong here. It should be ``... simulations and calculations of the AQN heat emission profile in the Earth's interior and the axion-flux distribution passing through the Earth's surface''. We did not compute the flux spectrum here. Sorry for the confusion.}. 

\subsection{Annihilation of AQNs impacting the Earth}
\label{subsec:annihilation of AQNs impacting the Earth}

The equations of motion describing the AQNs impacting the Earth are derived in Ref. \cite{Lawson:2019cvy}. We give a brief overview here. 

The energy loss due to the collision of AQNs with the Earth can be expressed as follows \cite{DeRujula:1984axn}
\begin{equation}
\label{eq:dE ds}
\frac{\rmd E}{\rmd s}=\frac{1}{v}\frac{\rmd E}{\rmd t}=-\sigma\rho v^2\ ,
\end{equation}
where $E=mv^2/2$ is the kinetic energy, $s$ is the path length, $\rho$ is the density of the local environment, $v$ is the AQN velocity, and $\sigma$ is the effective cross section of the AQN. In case of AQN annihilation within the Earth, the surrounding material is very rigid. Therefore, the cross section is the geometrical cross section,
\begin{equation}
\label{eq:sigma}
\sigma\simeq\pi R^2\ ,
\end{equation}
where $R$ is the radius of the AQN. Equation (\ref{eq:sigma}) implies that all nuclei which are on the AQN's    path  will get annihilated while the  AQN traverses the Earth. We refer readers to Ref. \cite{Lawson:2019cvy} for technical details and justification of this  argument.  

To proceed, following from the definition, we can express the time derivative of $E$ in the form
\begin{equation}
\label{eq:dE dt}
\frac{\rmd E}{\rmd t}
=mv\frac{\rmd v}{\rmd t}+\frac{1}{2}v^2\frac{\rmd m}{\rmd t}
=mv\frac{\rmd v}{\rmd t}-\frac{1}{2}\sigma\rho v^3\ ,
\end{equation}
where in the last step, we utilize the conventional rate of mass loss \eqref{eq:dm_AQN}. Comparing Eqs. \eqref{eq:dE ds} and \eqref{eq:dE dt}, we conclude that the mass loss acts like a friction term in the equation of motion
\begin{equation}
\label{eq:m dv dt}
m\frac{\rmd v}{\rmd t}
=-\frac{1}{2}\sigma\rho v^2
\simeq-\frac{1}{2}\pi R^2\rho v^2\ .
\end{equation}

The complete dynamical equations of motion in vector form are:
\begin{subequations}
	\label{eq:dr dt and dv dt}
	\begin{equation}
	\label{eq:dr dt}
	\frac{\rmd \mathbf{r}}{\rmd t}=\mathbf{v}\ ;
	\quad r=|\mathbf{r}|\ ;
	\quad v=|\mathbf{v}|\ ;
	\end{equation}
	\begin{equation}
	\label{eq:3.1 AQN DEs_v}
	\frac{\rmd \mathbf{v}}{\rmd t}
	=-\frac{1}{2}\pi R^2\frac{\rho(r)}{m}v^2\mathbf{\hat{v}}
	-\frac{GM_{\rm eff}(r)}{r^2}\mathbf{\hat{r}}\ ,
	\end{equation}
\end{subequations} 
where $G$ is the gravitational constant, and we define
\begin{subequations}
	\label{eq:R and M_eff(r)}
	\begin{equation}
	\label{eq:R}
	R=\left(\frac{3m}{4\pi\rho_\mathrm{n}}\right)^{1/3}
	\simeq1.045\times10^{-13}B^{1/3}\,{\rm cm}\ ,
	\end{equation}
	\begin{equation}
	\label{eq:M_eff(r)}
	\begin{aligned}
	&M_{\rm eff}(r)
	=\sum_{i=1}^{j}\frac{4\pi}{3}\rho_i(r_i^3-r_{i-1}^3)
	+\frac{4\pi}{3}\rho_{j+1}(r^3-r_j^3),  \\
	&(r_j\leq r< r_{j+1},~~{\rm with~} j=1, 2... 5~
	{\rm and~} r_0\equiv0)\ .
	\end{aligned}
	\end{equation}
\end{subequations}
In Eq. \eqref{eq:R}, we adopt the parameters in \cite{Raza:2018gpb,Lawson:2019cvy}, in which $\rho_\mathrm{n}=3.5\times10^{17}$\,kg\,m$^{-3}$ is the nuclear density and $m=m_p B$ is the AQN mass. In Eq. \eqref{eq:M_eff(r)}, we approximate the local environmental density $\rho(r)$ as discrete step functions due to the discontinuous geological structure of the Earth. The labels $i,j=1, ..., 5$ correspond to layers, summarized in Table \ref{tab:labels for each layers}. The parameter $r_i$ is the radius of the start of the layer as measured from the center of the Earth. $\rho_i$ is the average density of the corresponding shell. We make the approximation that the density within each layer is uniform, and therefore take the mean value from density at the top/bottom of the shell as the local density. The data in Table \ref{tab:labels for each layers} is taken from Ref. \cite{Anderson:1989}.

\begin{table}[h]
	\captionsetup{justification=raggedright}
	\caption{Our model for the density structure of the Earth. We consider the Earth to be made from 5 distinct layers, each of which we model as a constant-density shell, with density $\rho_i$. We show the outer radius for each shell, $r_i$, as well as the thickness of the shell.} % title of Table
	\centering % used for centering table
	\begin{tabular}{ccccc}
		\hline\hline
		Label & Layer   & Thickness [km] & $r_i$ [km]	 & $\rho_i$ [$\rm g\,cm^{-3}$]     \\
		\hline
		1     & Inner core  &1221  	& 1221	 &  12.95			\\
		2     & Outer core 	& 2259 &3480 	 & 11.05			\\
		3     & Lower mantle  & 2171 &5651	 & ~5.00			\\
		4     & Upper mantle   & \, 720 &6371	 & ~3.90			\\
		5     & Crust 	& \,\,~~30 & 6401
		& ~2.55 		    \\ 
		\hline\hline
	\end{tabular}
	\label{tab:labels for each layers}
\end{table}

\subsection{Axion-emission spectrum in the observer frame}
\label{subsec:axion-emission spectrum in laboratory frame}
We assume that the emission of AQN-induced axions is relativistic with velocity $v_a\sim0.6c$ and dominantly spherically symmetric in the AQN frame \cite{Fischer:2018niu,Liang:2018ecs}. In fact, spherical symmetry is well preserved even in the observer frame due to the relatively slow speed of AQNs $v_{\rm AQN}\sim 10^{-3}c\ll v_a$. To demonstrate this, we analyze the modification in the angular distribution due to the frame change. 

In general, an annihilating AQN will emit axions in a frame moving with respect to an observer on Earth. We introduce the notation $\tilde{K}$ and $K$ for the rest frame of the AQN and the frame of the observer respectively. The axion-emission velocity spectrum in the AQN rest frame is calculated in Ref. \cite{Liang:2018ecs} and we refer the reader to that paper for the calculation details. The velocity spectrum in the rest frame is, by definition, the derivative of the radiated flux $\Phi_{\rm rad}(\tilde{v}_a)$:
\begin{equation}
\label{eq:rho_rest(v_a)}
\begin{aligned}
\rho_{\rm rest}(\tilde{v}_a)
&\equiv
\frac{1}{\Phi_{\rm rad}^{\rm tot}}\frac{\rmd}{\rmd\tilde{v}_a}
\Phi_{\rm rad}(\tilde{v}_a)  \\
&\simeq\frac{\tilde{v}_a^3}{N(\delta)}
\left(\frac{\tilde{E}_a}{m_a}\right)^6|H_0(\tilde{p},\delta)|^2,
\end{aligned}
\end{equation}
where $\tilde{v}_a$ and $\tilde{E}_a$ are the rest frame axion velocity and energy respectively. The function $H_l(\tilde{p},\delta)$ corresponds to partial wave expansion in the approximate solutions, as derived in \cite{Liang:2018ecs}. As claimed in the beginning, annihilation of AQN is assumed to preserve spherical symmetry in the AQN rest frame, therefore only the $l=0$ mode is considered in Eq. \eqref{eq:rho_rest(v_a)}. The parameter $\delta\in (0,1)$ is a convenient factor introduced in Ref. \cite{Liang:2018ecs} as a result of approximations due to absence of simple analytic solutions of the general expressions. Tuning $\delta\in (0,1)$ leads to changes in the velocity spectrum \eqref{eq:rho_rest(v_a)} that do not exceed $\sim20\%$. The velocity spectrum is not known to better precision (see Appendix \ref{app:spectral properties in the rest frame}). The normalization factor  $N(\delta)$, depending on the parameter $\delta$, is also known and presented in Appendix \ref{app:spectral properties in the rest frame}.

In the frame of the observer, an AQN is moving with a velocity $v_{\rm AQN}\lesssim10^{-3}c$. Thus, we need only to consider a non-relativistic transformation of frames, with relations as follows: 
\begin{subequations}
	\label{eq:frame transformation}
	\begin{equation}
	\label{eq:frame transformation_vp}
	\mathbf{\tilde{v}}_a=\mathbf{v}_a-\mathbf{v}_{\rm AQN}\ ;
	\quad\mathbf{\tilde{p}}=\mathbf{p}-m_a\mathbf{v}_{\rm AQN}\ ;
	\end{equation}
	\begin{equation}
	\label{eq:frame transformation_v}
	\tilde{v}_a
	=\sqrt{v_a^2+v_{\rm AQN}^2-2\mathbf{v}_a\cdot\mathbf{v}_{\rm AQN}}
	\simeq v_a+{\cal O}(\frac{v_{\rm AQN}}{v_a})\ ;
	\end{equation}
	\begin{equation}
	\label{eq:frame transformation_p}
	\tilde{p}
	=\sqrt{p^2+m_a^2v_{\rm AQN}^2
		-2m_a\mathbf{p}\cdot\mathbf{v}_{\rm AQN}}\ ;
	\end{equation}
	\begin{equation}
	\label{eq:frame transformation_E}
	\tilde{E}_a
	=\sqrt{E_a^2
		+m_a^2v_{\rm AQN}^2
		-2m_a\mathbf{p}\cdot\mathbf{v}_{\rm AQN}}\ .
	\end{equation}
\end{subequations}
Working in the manifold of $\mathbf{v}_a$, we know $\rho(v_a)$ is normalized within a unit 3-ball $B^3$
\begin{equation}
\label{eq:rho transform}
1
=\int_{B^3} \rmd^3\mathbf{\tilde{v}}_a
\frac{\rho_{\rm rest}(\tilde{v}_a)}{4\pi \tilde{v}_a^2}
=\int_{B^3} \rmd^3\mathbf{v}_a
\frac{\rho_{\rm rest}[\tilde{v}_a(v_a)]}{4\pi (\tilde{v}_a(v_a))^2}\ ,
\end{equation}
where in the second step, we transform our coordinate from $\tilde{v}_a$ to $v_a$ using Eqs. \eqref{eq:frame transformation}. Note that such transformation produces a small error   related to a slight shift of spherical center by $\sim10^{-3}c$. However, this inconsistency is negligible comparing to the uncertainty of approximation in terms of $\delta$. Noting that the last equality in Eq. \eqref{eq:rho transform} is completely expressed in terms of $v_a$, the velocity in the frame of the observer. Therefore, we read off the emission spectrum in the frame of the observer
\begin{equation}
\label{eq:rho_obs(v_a)}
\rho_{\rm obs}(v_a)
=\int \rmd\Omega~v_a^2
\frac{\rho_{\rm rest}(\tilde{v}_a)}{4\pi \tilde{v}_a^2}\ ,
\end{equation} 
where $\Omega$ is the solid angle made by $\langle\hat{\mathbf{v}}_a,\hat{\mathbf{v}}_{\rm AQN}\rangle$. From Eq. \eqref{eq:frame transformation_v}, we expand Eq. \eqref{eq:rho_obs(v_a)} as
\begin{equation}
\label{eq:rho_obs(v_a) 2}
\begin{aligned}
\rho_{\rm obs}(v_a)
&\simeq\frac{v_a^3}{N(\delta)}\left(\frac{E_a}{m_a}\right)^6
|H_0(p,\delta)|+{\cal O}(\frac{v_{\rm AQN}}{v_a})  \\
&\simeq\rho_{\rm rest}(v_a)\ ,
\end{aligned}
\end{equation}
where ${\cal O}(\frac{v_{\rm AQN}}{v_a})\sim0.1\%$, the correction is negligible (comparing to the uncertainty of $\delta$) as claimed at the beginning of this subsection.

To summarize: the emission spectrum in frame of the observer \eqref{eq:rho_obs(v_a) 2} has an identical form, up to negligible correction, to the spherically symmetric spectrum in the rest frame \eqref{eq:rho_rest(v_a)} and computed in \cite{Liang:2018ecs}. This is expressed in terms $H_0(p,\delta)$ given in Appendix \ref{app:spectral properties in the rest frame}. One important fact to emphasize is, although spherical symmetry is well preserved in velocity spectrum of axion emission, the symmetry does not extend to the flux profile on Earth's surface due to asymmetric effects such as the daily modulation (see Sec. \ref{subsec:daily modulation}). In the next subsection, we introduce azimuthally symmetric function $P_a(\theta)$ to account for such effect.

\subsection{The axion flux density on Earth's surface}
\label{subsec:the axion flux density on Earth's surface}
The trajectory of an axion can be approximated as free motion because the gravity of the Earth is   too weak to modify the relativistic axions: $v_a\sim 0.6c\gg v_{\rm esc}=11\kmps$. The axion flux density on the surface of Earth is
\begin{equation}
\label{eq:Phi_a(theta)}
\Phi_a(\theta)
=\frac{\langle v_a\rangle}{c}\langle\dot{N}\rangle\langle N_a\rangle
\frac{P_a(\theta)}{2\pi R_\oplus^2}\ ,
\end{equation} 
where $\langle v_a\rangle\simeq0.6c$ is the average speed of emitted axion flux (see Appendix \ref{app:spectral properties in the rest frame} for numerical estimation), $\langle \dot{N}\rangle$ is the total hit rate of AQNs to surface of Earth, $\langle N_a\rangle$ is the total number of axions emitted per AQN, and $P_a(\theta)$ is the azimuthal distribution of the axion flux due to daily modulation (see Sec. \ref{subsec:daily modulation}). The first quantity is already estimated in \cite{Lawson:2019cvy}:
\begin{equation}
\label{eq:__dot(N)__}
\langle\dot{N}\rangle
%\simeq2.12\times10^{7}{\rm yr^{-1}}
%\left(\frac{\rho_{\rm DM}}{0.3{\rm \frac{GeV}{cm^3}}}\right)
%\left(\frac{\mu}{220\kmps}\right)
%\left(\frac{10^{25}}{\langle B\rangle}\right)
=0.672 {\rm\, s}^{-1}
\left(\frac{\mu}{220\kmps}\right)
\left(\frac{10^{25}}{\langle B\rangle}\right)\ .
\end{equation}
The second quantity is proportional to the average mass loss per AQN $\langle \Delta m_{\rm AQN}\rangle$:
\begin{equation}
\label{eq:__N_a__}
\langle N_a\rangle
\simeq\frac{1}{3}\frac{\langle\Delta m_{\rm AQN}\rangle\,c^2}{\langle E_a\rangle}
\simeq\frac{\langle\Delta m_{\rm AQN}\rangle\,c^2}{4m_a}\ ,
\end{equation}
where $\langle E_a\rangle\simeq1.3m_a c^2$ is the average energy of axion emitted given by the spectrum \eqref{eq:rho_obs(v_a) 2}, also see Ref. \cite{Liang:2018ecs}. The last quantity assumes an azimuthal symmetry of system and is defined by the normalization condition:
\begin{equation}
\label{eq:P_a}
1=\int_0^\pi\rmd\theta\sin\theta\cdot P_a(\theta)\ .
\end{equation}
Lastly, the energy flux on Earth's surface is
\begin{equation}
\label{eq:E_a Phi_a(theta)}
\langle E_a\rangle\Phi_a(\theta)
=\frac{\langle v_a\rangle\langle\dot{N}\rangle\langle \Delta m_{\rm AQN}\rangle\,c}{6\pi R_\oplus^2}P_a(\theta)\,.
%=\frac{1}{5}\langle\dot{N}\rangle\langle \Delta m_{\rm AQN}\rangle
%\frac{P_a(\theta)}{2\pi R_\oplus^2}
\end{equation}
Note that both the energy flux and energy density of the AQN-induced axions is independent of axion mass $m_a$, as the $m_a$ dependence in $\langle E_a\rangle$ and $\langle N_a\rangle$ cancels. This is a distinct feature comparing to conventional galactic axions which are $m_a$-sensitive.

\section{algorithm and simulation}
\label{sec:algorithm and simulation}

\subsection{Simulating initial conditions of AQNs for the heat emission profile}
\label{subsec:simulating initial conditions of AQNs for the heat emission profile}

\begin{figure}[h]
	\centering
	\captionsetup{justification=raggedright}
	\captionsetup{justification=raggedright}
	\includegraphics[width=0.7\linewidth]{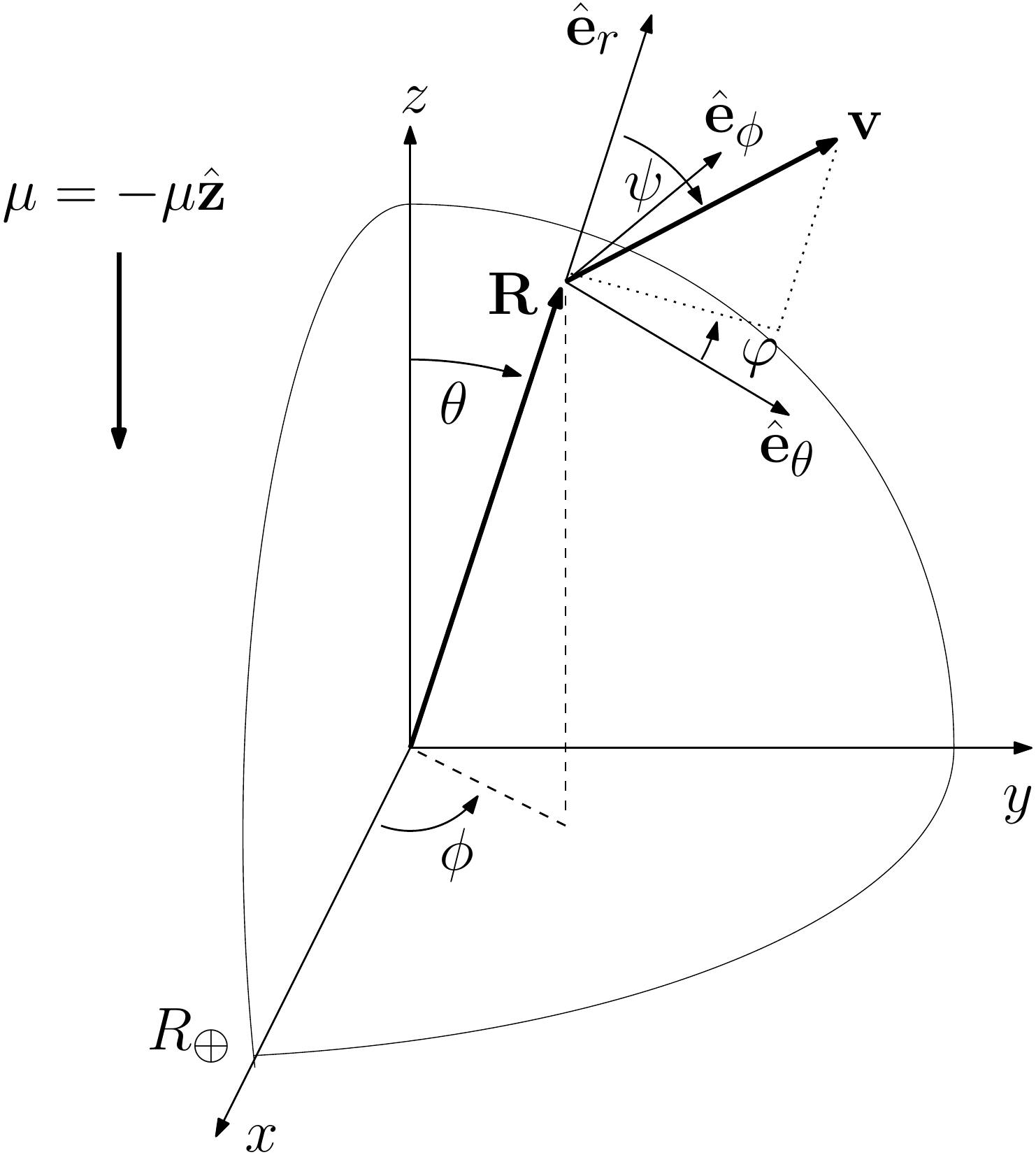}
	\caption{Coordinate system used in flux distribution of the AQN wind.}
	\label{fig:FluxCoord}
\end{figure}

The flux of relativistic axions is emitted instantaneously once an AQN looses mass through baryonic annihilation with matter on Earth. As shown in Fig. \ref{fig:FluxCoord}, to simulate the trajectories of AQNs through Earth we use the \textit{flux} distribution function of AQNs in the vicinity of Earth with wind coming at fixed direction \cite{Lawson:2019cvy}:
\begin{equation}
\label{eq:d dot(N) dv}
\begin{aligned}
\frac{\rmd}{\rmd v}\dot{N}
&=C\int_0^{\pi}\rmd\theta\int_{\frac{\pi}{2}}^\pi\rmd\psi
\int_0^{2\pi}\rmd\varphi\, v^3 e^{-\frac{v^2}{2\sigma^2}}\sin\theta\sin\psi\cos\psi \\
&\quad\times\exp\left[
-\frac{v\mu}{\sigma^2}(\cos\psi\cos\theta-\sin\psi\cos\varphi\sin\theta)
\right]\ ,
\end{aligned}
\end{equation}
where $C$ is a normalization constant, $\mu$ is the mean speed of AQN wind as defined in \eqref{eq:mu(t)}, the notation of angles ($\theta$, $\psi$, and $\varphi$) is presented in Fig. \ref{fig:FluxCoord}. 

To generate AQN initial conditions, we pick a point on the Earth's surface determined by the flux distribution function - the integrand of \eqref{eq:d dot(N) dv}. We sample initial conditions $v$, $\theta$,  $\psi$ and $\varphi$ from the function defined on a 4D space of $v$, $\theta$,  $\psi$ and $\varphi$. The function has no dependence on $\phi$ due to our approximation of the Earth as a sphere, so we set it to 0 for our simulation. We remove AQNs with initial velocities pointing away from the Earth as these correspond to AQNs that would have already experienced an Earth interaction. Following this sampling scheme, we generate $2\times10^5$ AQNs and compute $q(r,\theta)$, the heat emission profile of axions at a given location within the Earth $(r,\theta)$, subject to the following normalization condition 
\begin{equation}
\label{eq:q(r,theta)}
\langle \Delta m_{\rm AQN}\rangle
=\int_0^{R_{\oplus}}dr\int_0^\pi d\theta~q(r,\theta)\ ,
\end{equation}
$\theta$ is the angle between $\mathbf{r}$ and the (opposite-direction) galactic wind $\boldsymbol{\mu}$, see Fig. \ref{fig:FluxCoord}. Note that the normalization condition of $q(r,\theta)$ is not defined in a standard way where the solid angle term $\sin\theta$ is implicitly absorbed into $q(r,\theta)$ for simplicity.

\begin{figure}
	\centering
	\captionsetup{justification=raggedright}
	\includegraphics[width=0.7\linewidth]{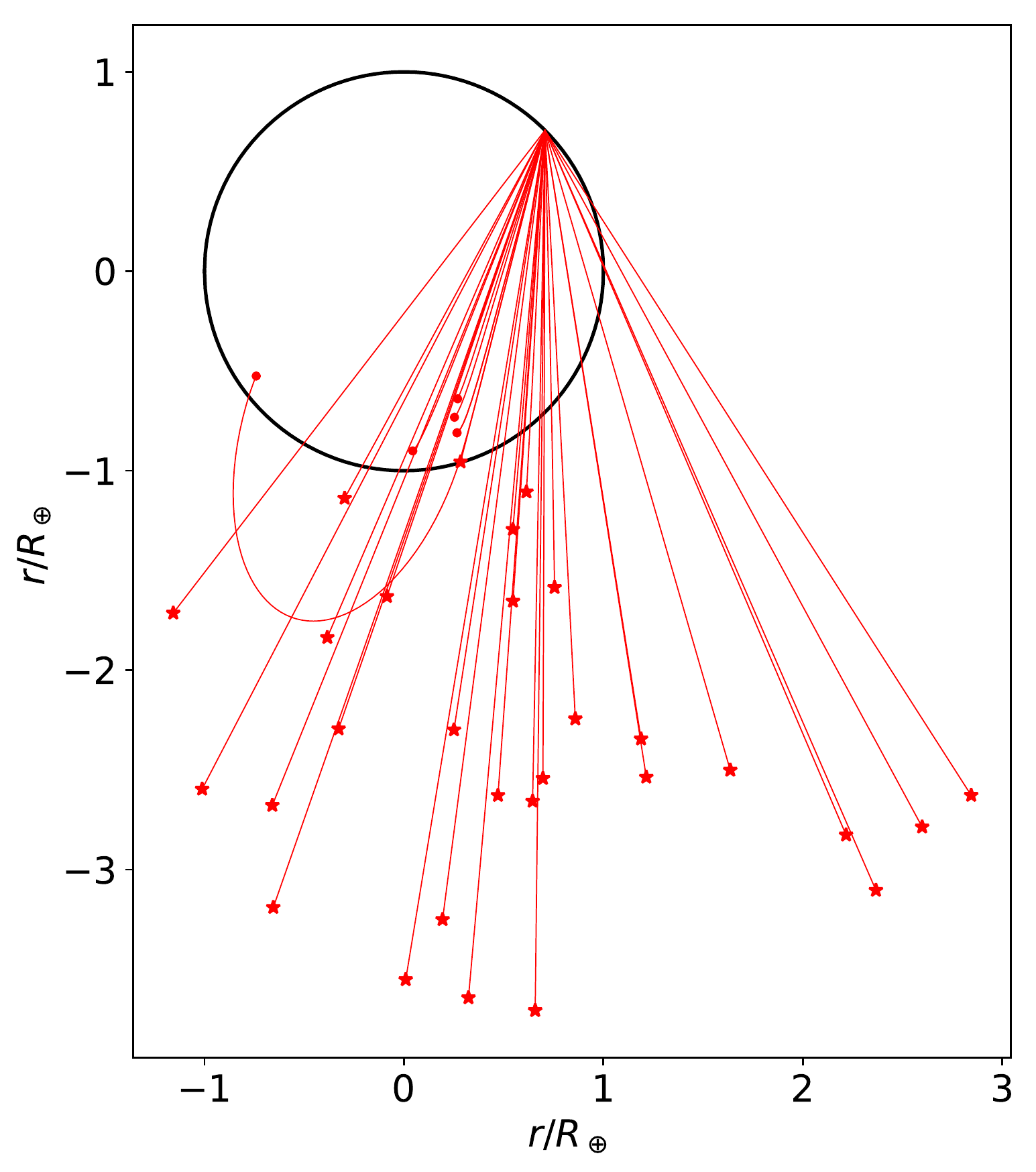}
	\caption{Some sample trajectories of AQNs (shown here to emanate from a single point on the Earth's surface for aesthetic purposes). The trajectories that have a star symbol at the end escape, while those that end in dots have been trapped. Most AQNs pass through the Earth and escape, whilst those that get trapped either never exit the Earth after entering, or leave the Earth and circle back into the Earth (or circle a few times, losing mass and velocity each time as they go through the Earth) and annihilate inside. The fraction of AQNs that are captured, and end up annihilating inside the Earth is dependent on the model $(\alpha,B_{\rm min})$, however the vast majority of AQNs escape.} 
	\label{fig:AQN trajectories}
\end{figure}

We also present a few AQN trajectories in Fig. \ref{fig:AQN trajectories}. In the plot the AQNs are seen to be entering the Earth from one point on the Earth's surface for plotting purposes. However, the initial entering velocities of the AQNs, and their positions on the Earth's surface are determined by the flux distribution function and have been accounted for in our simulations. Their trajectories are determined by Eqs. \eqref{eq:dr dt and dv dt}. As expected, the vast majority of AQNs penetrate and escape from the Earth in linear motions due to the weakness of gravitation compared to the kinetic energy of AQNs ($v_{\rm esc}/v_{\rm AQN}\sim0.1$). We also present a few trajectories of trapped AQNs (up to negligible portion) for illustrative purpose.

\subsection{Computations of the surface axion flux density}
\label{subsec:computations of the surface axion flux density}
As shown in Eq. \eqref{eq:E_a Phi_a(theta)} the magnitude of axion flux on the surface of Earth is determined by the mass loss $\langle m_{\rm AQN}\rangle$ simulated in the preceding subsection. However, the azimuthal component is directly related to $P_a(\theta)$, the surface probability of finding an axion at a given azimuthal angle $\theta$, given by Eq. \eqref{eq:P_a}. To obtain $P_a(\theta)$, we use Monte Carlo simulations based on the heat profile $q(r,\theta)$. First, it is clear that the number of axions to simulate at a given point at position $(r,\theta)$ is proportional to the rate of axion production [i.e. the heat emission profile function $q(r,\theta)$]. Therefore, we sample axion initial conditions from the distribution within the integral
\begin{equation}
\label{eq:P_a(theta)}
\begin{aligned}
P_a(\theta)
\sim\langle N_a\rangle  
\propto\int_0^{R_{\oplus}}\rmd r\int_{0}^{\pi}\rmd \theta~
\frac{q(r,\theta)}{4m_a} \ .
\end{aligned}
\end{equation}

\begin{figure}[h]
	\centering
	\captionsetup{justification=raggedright}
	\includegraphics[width=0.7\linewidth]{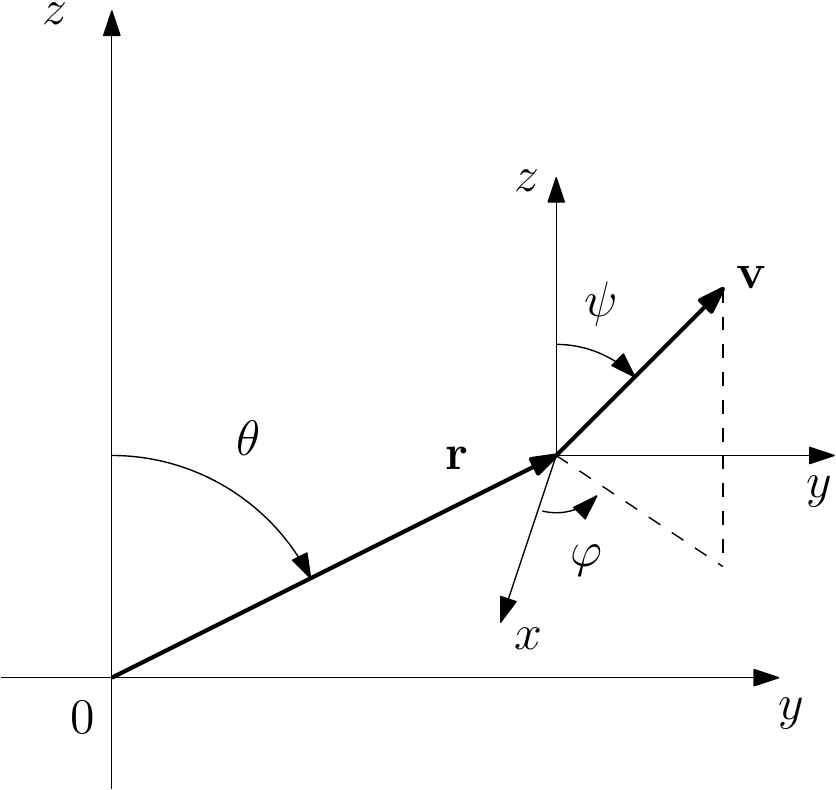}
	\caption{Monte Carlo simulation of axion flux. The initial position is chosen on the $y$-$z$ plane (by azimuthal symmetry) drawn from heat emission profile $q(r,\theta)$. Then an axion is launched at a random 3D solid angle $(\psi,\varphi)$, and with magnitude $v_a$.}
	\label{fig:axion simulate}
\end{figure}

As shown in Fig. \ref{fig:axion simulate}, we choose the initial position vector of emitted axion to be on the $y$-$z$ plane by azimuthal symmetry of $q(r,\theta)$. In addition, we know gravitation of Earth is negligible because axion has relativistic speed $v_a\sim0.6c$ and largely exceeds the escape velocity $\sim10^{-5}c$. Therefore given an emitted axion starting at $(r_0,\theta_0)$, its trajectory is simply a straight line of uniform motion
\begin{subequations}
\label{eq:r and v0, r0}
\begin{equation}
\label{eq:r(t)}
\mathbf{r}(t)
=\mathbf{v}t+\mathbf{r}_0
\equiv v_at\hat{\mathbf{v}}+r_0\hat{\mathbf{r}}_0\ ,
\end{equation}
\begin{equation}
\label{eq:v0, r0}
\hat{\mathbf{v}}
=(\sin\psi\cos\varphi,\sin\psi\sin\varphi,\cos\psi),\,
\hat{\mathbf{r}}_0
=(0,\sin\theta_0,\cos\theta_0),
\end{equation}
\end{subequations}
where the initial velocity vector is sampling as follows:
\begin{equation}
\label{eq:v_a, cos psi, varphi}
\begin{aligned}
v_a\sim\rho_{\rm obs}(v_a);~
\cos\psi\sim{\rm Unif}[-1,1];~
\varphi\sim{\rm Unif}[0,2\pi]\, .
\end{aligned}
\end{equation}
Here `Unif' stands for uniform distribution. Note that the equation of motion \eqref{eq:r and v0, r0} gives the intercept point(s) at the Earth surface $r=R_\oplus$ and hence the  angle $\theta$ in the azimuthal distribution $P_a(\theta)$. Therefore solving for
\begin{equation}
\label{eq:R_oplus2}
R_\oplus^2=\mathbf{r}^2\ ,
\end{equation}
we find the angle $\theta$ on surface of the Earth:
\begin{equation}
\label{eq:R_oplus cos(theta)}
\begin{aligned}
\cos\theta
&=\frac{r_0}{R_\oplus}\cos\theta_0  \\
&\quad-\frac{r_0}{R_\oplus}\cos\psi\left[
\hat{\mathbf{v}}\cdot\hat{\mathbf{r}}_0
\pm\sqrt{(\hat{\mathbf{v}}\cdot\hat{\mathbf{r}}_0)^2
	+\left(\frac{R_\oplus^2}{r^2}-1\right)}
\right]\, .
\end{aligned}
\end{equation}
Note that the solution gives two roots in Eq. \eqref{eq:R_oplus2}: one positive and one negative solution of $t$ in Eq. \eqref{eq:r and v0, r0}. By simulating sufficient numbers of axions based on initial conditions \eqref{eq:r and v0, r0} and \eqref{eq:v_a, cos psi, varphi}, we obtain the the azimuthal distribution $P_a(\theta)$.

As an additional note, the simulation is not efficient by picking out the positive-time solution only each time from the sampling array. In practice, we do not distinguish the two solutions and put both equally into statistics. In other words, we launch a pair of axions, instead of one only, with opposite initial velocity at position $(r_0,\theta_0)$ in the simulation code.

\section{Discussion of results}
\label{sec:discussion of results}

\subsection{Heat emission profile}
\label{subsec:heat emission profile}
We first discuss  the results of the simulation ($2\times10^5$ samples) for the heat-emission profiles of axions $q(r,\theta)$. In our simulations we use a number of parameters such as the AQN baryon-charge distribution (parameters $\alpha$ and $B_{\rm min}$) and flux distribution of incoming AQNs (e.g. the annual/daily modulation and solar gravitation). Surprisingly, we find that our results are quite insensitive to these parameters. Therefore, in the main body of the paper we only present the case with $(\alpha,B_{\rm min})=(2.5,3\times10^{24})$ shown in
Fig. \ref{fig:q(r,theta)_alpha25}, while leaving other cases with different parameters to Appendix \ref{app:on sensitivity of the main results to the AQN's parameters}. 
From Fig. \ref{fig:q(r,theta)_alpha25}, we observe the profile is in general as like a typical collection set of linearly moving trajectories in polar coordinate.  This is as expected because the gravitational force is weak compared to the kinetic energy of AQNs.

\exclude{Our next comment goes as follows.} The heat emission is greater near the upper pole $\theta=0$ with respect to the wind direction, despite of the fact that every annihilation event  is assumed to be spherically symmetric. This is the cause of daily modulation explained in Sec. \ref{subsec:daily modulation}: the AQN emits more heat when it enters Earth compared to when it leaves due to a larger initial cross section. 

In addition, we observe abrupt changes of heat emission at some specific distance (e.g. $r\sim0.9R_\oplus$ and $0.5R_\oplus$). One can see, from Table \ref{tab:labels for each layers}, that these jumps precisely correspond to the successive layers of the Earth interior. Whenever an AQN moves into a new layer with an abrupt change of local density, the mass loss drastically changes.

Another observation is that very few AQNs reach the core of the Earth, as the dominant portion of the AQNs propagate at the distances $r\geq0.5R_\oplus$. This is a geometrical effect that stems from the fact that
%there is more volume in the outer part of the Earth compared to the inner part, and 
very few AQNs hit the surface with zero incident angle.  

In Appendix \ref{app:on sensitivity of the main results to the AQN's parameters} we study the
sensitivity of our results to the parameters of the system.
We observe that   almost all the effects discussed in the paper are insensitive to specific choices for the parameters of the model such as    the baryon-charge distribution of AQN models and   the velocity distribution of incoming DM flux. Numerical simulations confirm this behavior.

\begin{figure}[h]
	\centering
	\captionsetup{justification=raggedright}
	\includegraphics[width=\linewidth]{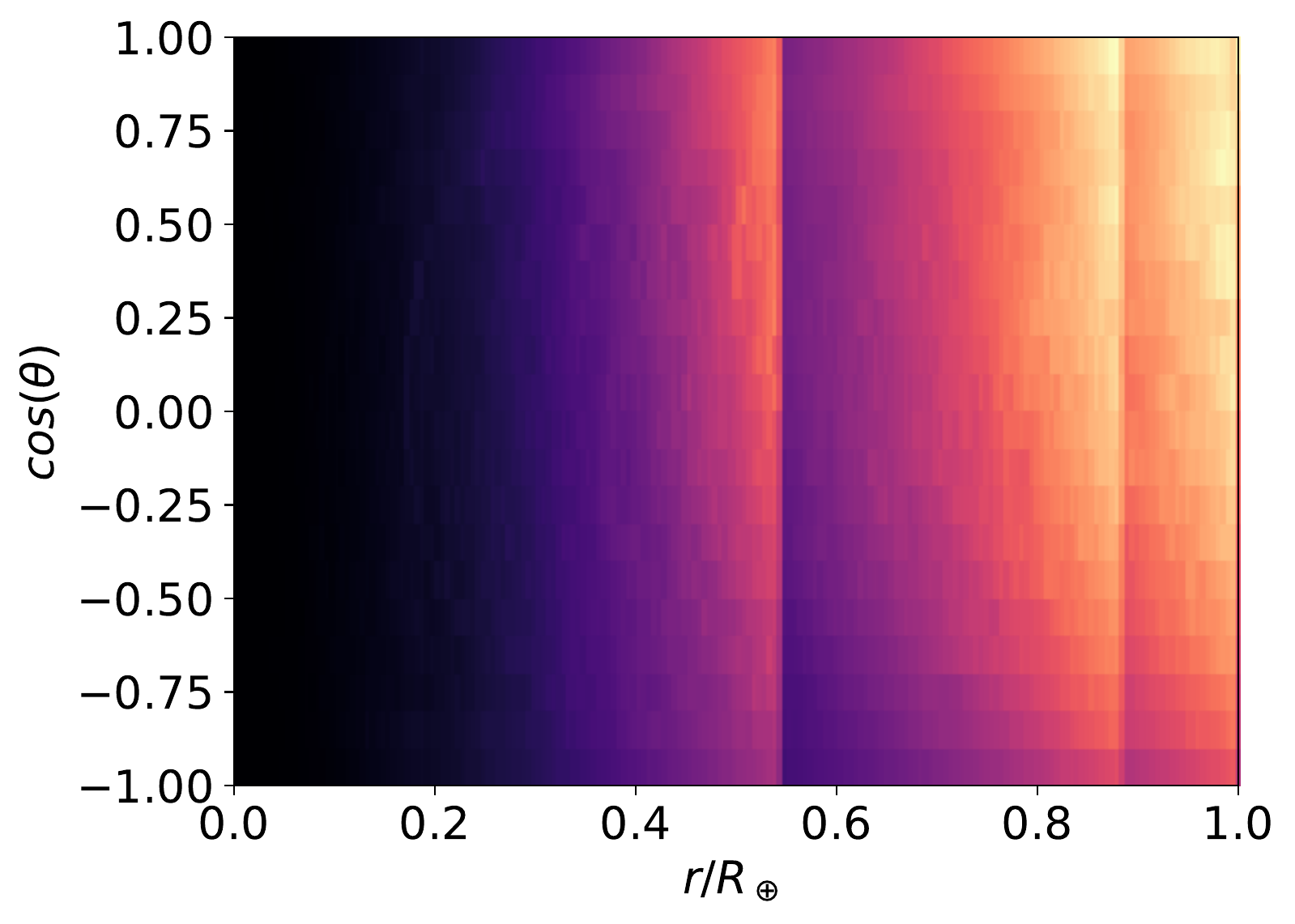}
	\caption{Probability density for the heat-emission profile of axions within the Earth: $q(r,\theta)$ for $(\alpha,B_{\rm min})=(2.5,3\times10^{24})$. The coordinates are with respect to the DM wind direction at $\theta=0$. The profile is very similar to a typical collection set of linearly moving trajectories in polar coordinate because of the weakness of gravitational effect comparing to kinetic momentum. The azimuthal asymmetry corresponds to the daily modulation explained in Sec. \ref{subsec:daily modulation}. The abrupt jumps at some specific radii are due to discontinuous change of local density between successive layers of the Earth interior. $2\times10^5$ samples were used for the simulation.}
	\label{fig:q(r,theta)_alpha25}
\end{figure}

\subsection{Axion flux density}
\label{subsec:axion flux density}
The key results of our simulations are summarized in Table \ref{tab:summary of some results}. The axion flux $\langle E_a\rangle\Phi_a$ is obtianed from Eq. \eqref{eq:E_a Phi_a(theta)} and the energy density $\langle\rho_a\rangle$ is the axion flux divided by $\langle v_a\rangle\simeq0.6c$. We also present the azimuthal distribution $P_a(\theta)$ as defined in Eq. \eqref{eq:P_a} for $(\alpha,B_{\rm min})=(2.5,3\times10^{24})$. Similar to the conclusion in the preceding subsection, we find that the results are not sensitive to the modeling parameters: the axion flux $\langle E_a\rangle\Phi_a$ in general constrains within $(10^{13}-10^{14})\,\rm eV\,cm^{-2}s^{-1}$ depending on the size distribution, similar result applies for the density $\langle\rho_a\rangle\sim(10^{-6}-10^{-5})\,\rm GeV\,cm^{-3}$. While the main effect of modulations and enhancements on the axion flux have been discussed in Section \ref{sec:potential enhancements}, here we want to make few additional comments on these results (we refer the reader to Appendix \ref{app:on sensitivity of the main results to the AQN's parameters} for further technical details on the sensitivity to the parameters of the system).

Regarding the annual modulation, we explored two extreme cases of mean galactic velocity $\mu$ with $\mu=V_\odot\pm V_\oplus$ [see Eq. \eqref{eq:mu(t)} and Table \ref{tab:summary of some results}]. We find the flux is modulated by ${\cal O}(10\%)$ as expected from discussion in Sec. \ref{subsec:annual modulation}. Regarding the daily modulations, the results are shown in Table \ref{tab:summary of some results} as the fluctuations $\pm$ in the last column.  We find that the magnitude of daily modulation is of order ${\cal O}(10\%)$ and proportional to the ratio of average mass loss $\langle \Delta B\rangle/\langle B\rangle$, which is
  consistent with our simplified  estimate  \eqref{eq:percetage fluctuation}.

  We also studied   the related effect on the azimuthal distribution $P_a(\theta)$ plotted on Fig. \ref{fig:P_a(r,theta)_alpha25}.   It shows nearly linear dependence on $\cos\theta$ which is consistent
  with our interpretation because  the mass loss is indeed proportional to the path length $s$ and therefore to $\cos\theta$, according to Eq. \eqref{eq:dm_AQN}.
  %As shown in Fig. \ref{fig:P_a(r,theta)_alpha25}, we verify such linear proportionality holds well for $P_a(r,\theta)$ as a function of $\cos\theta$ [noting that the mass loss is proportional to path length $s$ and therefore $\cos\theta$, as argued in Eq. \eqref{eq:dm_AQN}]. 
  We also note there is a noticeable jump near the upside pole ($\cos\theta=1$). This is interpreted as a sharp impact caused by the initial stage of the annihilation process. 

Finally, it is instructive to compare the result from Table \ref{tab:summary of some results} to the order of magnitude estimate \eqref{eq:rho_a intro} presented in Ref.  \cite{Fischer:2018niu}. There is approximately two orders of magnitude deviation in numerical factor from that naive estimate. The reason for the discrepancy can be understood as follow: First, the estimate in Ref. \cite{Fischer:2018niu} assumed that $\Delta B/B\sim 1$ (i.e., most incident AQN are completely  annihilated). However, our simulations show that $\Delta B/B\sim 0.1$ in most cases, see Table \ref{tab:summary of some results}. Similarly there is a numerical factor of 3/5 that is neglected in Ref. \cite{Fischer:2018niu} as only the AQNs made out of antiquarks will be annihilated underground. Finally, there is a geometrical suppression factor related to averaging over inclination angles between the AQN velocity and the surface of the Earth that was ignored  in  \cite{Fischer:2018niu}.

\begin{table*} [!htp]%[h] here; [t] top; [b] bottom
	\caption{Summary of some results ($B_{\rm min}=3\times10^{24}$, $\epsilon=1$ unless specified, $10^8$ samples), the uncertainties in axion flux $\langle E_a\rangle\Phi_a$ and density $\langle \rho_a\rangle$ describe the maximum daily modulation. Note that the AQN-induced energy flux and energy density are independent of axion mass in contrast to the conventional galactic axion models, see discussion at the end of Sec. \ref{subsec:the axion flux density on Earth's surface}.} % title of Table
	\centering % used for centering table
	\captionsetup{justification=raggedright}
	\begin{tabular}{ccccccc} % centered columns (4 columns)
		\hline\hline %inserts double horizontal lines
		$\langle B\rangle$ &$\alpha$ &Other parameters & $\langle \Delta m_{\rm AQN}\rangle$ [kg] & ${\langle\Delta B\rangle}/{\langle B\rangle}$ & $\langle E_a\rangle\Phi_a~\rm[10^{13}\frac{eV}{cm^2s}]$ 
		& $\langle\rho_a\rangle~\rm[10^{-6}\frac{GeV}{cm^{3}}]$ \\\hline
		$8.84\times10^{24}$ &2.5  &--      & $4.96\times10^{-3}$                 &  $33.5\% $                                       &  $8.12\pm0.93$   & $4.51\pm0.52$                                 \\
		$8.84\times10^{24}$ &2.5  &$\mu=V_\odot - V_\oplus$      & $4.93\times10^{-3}$                 &  $33.4\% $                                       &  $6.99\pm0.72$  & $3.88\pm0.40$                                \\
		$8.84\times10^{24}$ &2.5  &$\mu=V_\odot + V_\oplus$      & $5.00\times10^{-3}$                 &  $33.8\% $                                       &  $9.30\pm1.17$  & $5.17\pm0.65$             \\
%		$8.84\times10^{24}$ &2.5  &$\epsilon=3$      & $1.33\times10^{-2}$                 &  $90.0\% $                                       &  $(2.33\pm1.29)\times10^{14}$    &  75.1\%                                   \\
		$8.84\times10^{24}$ &2.5  &Solar gravitation\footnote{We consider an additional $42.1\kmps$ AQN impact velocity from AQN falling to Earth from infinity in the gravitational well of the Sun, this gives an additional velocity to the AQN that is always in the direction of travel. See Appendix \ref{app:on sensitivity of the main results to the AQN's parameters} for more details.}      & $4.94\times10^{-3}$                 &  $33.4\% $                                       &  $9.67\pm1.05$ & $5.37\pm0.58$             \\
		$2.43\times10^{25}$ &2.0  &--     & $8.16\times10^{-3}$                     & $20.1\%$                                         & $4.91\pm0.43$   & $2.73\pm0.24$          \\
		$4.25\times10^{25}$ &(1.2, 2.5) & $B_{\rm min}=10^{23}$       & $1.04\times10^{-2}$                &  $14.7\%$                                     & $3.60\pm0.20$   & $2.00\pm0.11$                   \\
		$1.05\times10^{26}$ &(1.2, 2.5) &--     & $2.48\times10^{-2}$                     & $14.1\%$                                         & $3.46\pm0.16$    & $1.92\pm0.09$                   \\\hline\hline
	\end{tabular}
	\label{tab:summary of some results} % is used to refer this table in the text
\end{table*}

\begin{figure}
	\centering
	\captionsetup{justification=raggedright}
	\includegraphics[width=\linewidth]{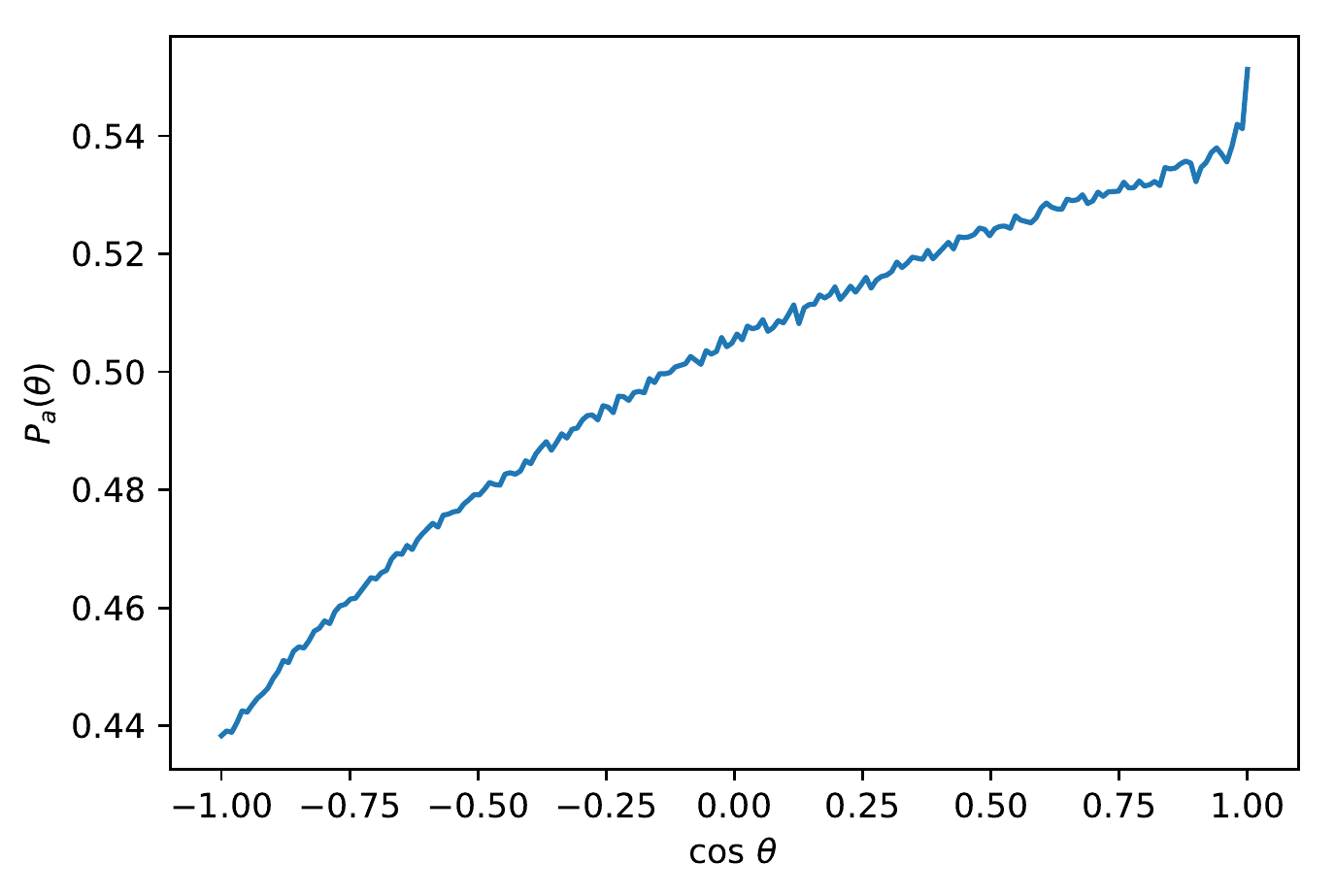}
	\caption{The azimuthal distribution of axion flux on the surface of the Earth: $P_a(\theta)$ for $(\alpha,B_{\rm min})=(2.5,3\times10^{24})$. The shape of $P_a(\theta)$ is in general a linear function of $\cos\theta$. The magnitude of asymmetry is well consistent with the estimation \eqref{eq:percetage fluctuation}. The existence of such azimuthal asymmetry is the key to daily modulation. $10^8$ samples were used for the simulation.}
	\label{fig:P_a(r,theta)_alpha25}
\end{figure}

\section{Conclusions and future directions}
\label{sec:conclusions and future directions}
The goal of the present work was to perform the  Monte Carlo simulations  describing the distributions of the AQN trajectories and annihilation events in the Earth's interior in order to calculate the AQN-induced axions with $\la v_a\ra\simeq 0.6c$. 
This allowed us  to compute a number of time dependent modulation and amplification effects which have been heavily used in \cite{Budker:2019zka}. Our goal was not to discuss the detection of these AQN-induced axions with a specific instrumentation, we refer to \cite{Budker:2019zka} where the detection issues were addressed.
The main results of the present work can be summarized as follow:\\

{\bf a}. We computed the   axion flux, the spectrum and the angular distribution  of the axions generated by antimatter AQNs crossing the Earth using  Monte Carlo simulations. These axions are produced when the antimatter AQNs annihilate with the rocky material in the Earth interior.   The main results are presented in Table \ref{tab:summary of some results}.

{\bf b}. We computed the  time-dependent the  annual and daily  modulations   in the axion intensity, see Table  \ref{tab:potential enhancement}. Some of the predicted effects are unique to the AQN framework and constitute a decisive test of the model. The corresponding results have been heavily  used in    accompanying paper \cite{Budker:2019zka} where the broadband detection strategy has been proposed.

{\bf c}. We also computed the  time-dependent burst like enhancement, which we coined as  the ``local flash".  This amplification  could be enormous $(10^2-10^4)$, see Table \ref{tab:local flashes estimation}.  This strong enhancement occurs  when AQN hits the earth close enough from the position of the axion search detector.  
 This effect has been heavily  used in    accompanying paper \cite{Budker:2019zka}   
 where a powerful test to discriminate  the  true axion signal from a spurious background  noise, has been suggested.

The effects discussed in items {\bf a-c}   depend very weakly on the parameters of the model such as the nuggets size-distribution, consequently, there is little to no flexibility in the predictions presented in Tables \ref{tab:potential enhancement}, \ref{tab:local flashes estimation} and  \ref{tab:summary of some results}.

Why should we consider the AQN model seriously? 
%The present study shows that the model makes specific predictions that future axion direct detection experiments should be able to see. The direct observation of relativistic axions, together with the gravitationally trapped axions studied in our previous work \cite{Lawson:2019cvy}, with their very distinct density and velocity-spectral properties would be the smoking gun supporting the entire AQN framework.
From an observational cosmology angle, this model is consistent with all available cosmological, astrophysical, satellite and ground based constraints, where AQNs could leave a detectable electromagnetic signature. It provides a simple explanation for the observed relation $\Omega_{\rm DM}\sim \Omega_{\rm visible}$ and the baryon asymmetry without the need for fundamental changes or extensions of the standard model. The baryogenesis is replaced by ``charge separation" effect. It was also showed that the AQNs   can form and  survive the very early epochs of the evolution of the Universe, such that  they may serve as the DM candidates today.   The same AQN framework may also explain a number of other (naively unrelated) phenomena, such as the  excess of galactic emission in different frequency bands; it may  also offer resolutions to some other astrophysical mysteries such as  ``Primordial Lithium Puzzle",   so-called ``The Solar Corona Mystery", the DAMA/LIBRA observed annual modulation, as well as  the recent EDGES observations of a stronger than anticipated 21 cm absorption features, see Introduction for references and details.

Interestingly, the results of the present work could also be used for different purposes, not directly   related to the axion searches, such as analysis of the  annual modulation in the AQN-induced neutrino intensity. Such a study    could be  a  key element in explanation \cite{Zhitnitsky:2019tbh} of the 20 years old DAMA/LIBRA puzzling observation of the annual modulation.

\section*{Acknowledgments}
This work was supported in part by the National Science and Engineering Research Council of Canada.
 AZ   thanks   Yannis Semertzidis  for explaining the role of different time scales (cavity storage time, axion coherence time etc) in axion search experiments.  AZ also thanks many participants of the workshop ``Axion Experiments in Germany", August 19-22, 2019 and the   ``IBS Conference on Dark World", Daejeon, Korea, November 4-7, 2019  where this work has been presented,  for discussions and large number of  good questions.   AM acknowledges support from the Horizon 2020 research and innovation programme of the European Union under the Marie Sk\l{}odowska-Curie grant agreement No. 702971. We thank the authors of ref. \cite{Bertolucci:2019jsd} for correspondence which resulted in   clarification of our claims in Section \ref{subsec:gravitational lensing}. 

\appendix
\section{Few comments on the broadband   detection strategy.}
\label{broadband}
The goal here is to highlight few ideas on   broadband detection  strategy as suggested in   \cite{Budker:2019zka}. 
First of all, we remark that 
the axion field $a(\mathbf{r}, t)$ can be treated as a classical field because the number   of the AQN-induced axions \eqref{eq:m_a Phi_a simeq}, \eqref{eq:E_a Phi_a numerical} accommodated by a single de-Broglie volume is very large in spite of the fact that the de-Broglie wavelength $\lambda$ for relativistic AQN-induced axions  is much shorter than for conventional galactic axions, 

 \be
\label{density}
n_a\lambda^3\sim \frac{\Phi_a}{v_a}\cdot \left(\frac{\hbar}{m_a v_a}\right)^3\sim 10^6 \left(\frac{10^{-4} {\rm eV}}{m_a}\right)^4\gg 1. \nonumber
\ee

\exclude{
\be
\label{density}
n_a\lambda^3\sim \frac{\rho_a}{m_a}\cdot \left(\frac{\hbar}{m_a v_a}\right)^3\sim 10^6 \left(\frac{10^{-4} {\rm eV}}{m_a}\right)^4\gg 1. \nonumber
\ee
}

We start our overview with QUAX \cite{Barbieri:2016vwg}, CASPEr \cite{JacksonKimball:2017elr} ideas when the basic coupling is the interaction between the gradient of the 
axion field and the spin, 
\be
\label{H}
H_{\rm spin}\simeq g_{\rm a} \boldsymbol{\sigma}\cdot \boldsymbol{\nabla} a(\mathbf{r}, t),~~~ g_{\rm a }\propto f_a^{-1}.
\ee
In formula (\ref{H})   we use, for generality,  a single parameter $g_a$  which may assume the value   $g_{\rm a }\equiv g_{\rm aee}$ for electrons in case of QUAX  or $g_{\rm a }\equiv g_{\rm aNN}$ for nucleons in case of CASPEr.  The coupling is proportional  to 
the axion  velocity $\mathbf{v}_a \propto\boldsymbol{\nabla} a(\mathbf{r}, t)$ such that it is 3 orders of magnitude enhancement 
for the AQN axions with $v_a\sim 0.6 c$ in comparison with galactic axions with $v_a\sim 10^{-3}c$.  For the coupling (\ref{H}) one can  use a broadband detection  technique if a single photon detectors  for GHz range  become available, which is claimed to be the case \cite{Kuzmin2013,Lamoreaux:2013koa}.

There is another type of instruments such as ABRACADABRA \cite{Kahn:2016aff}, LC Circuit \cite{Sikivie:2013laa}, DM Radio \cite{Chaudhuri:2018rqn} and others which can operate in the resonance as well as  in broadband regimes. A similar design   \cite{Cao:2017ocv}
was  also suggested as an instrument capable to measure the intensity of axion field $ {a}(\mathbf{r}, t)$ by using the Topological   Casimir Effect.  All these designs are based on the idea that the time-dependent axion field generates an additional electric current $\mathbf{j}_a \propto  \dot{a}(\mathbf{r}, t)\,\mathbf{B}$ in the presence of the background magnetic field $\mathbf{B}$. All the instruments which are mentioned  above are very different in sizes and designs. However, their common feature is that  these instruments are capable to measure   the induced electric current in pickup loop \cite{Kahn:2016aff}   which is obviously represents a broadband capability  not shared by the   cavity type  instruments.

We  reiterate the main  essence of the proposal.  The conventional cavity detection technique search assumes a scanning  to be   sensitive to  very narrow   resonance line. The  broadband strategy \cite{Budker:2019zka} can be used to probe the AQN emitted  axions using the daily or annual modulation computed in the present work to select some specific frequency bin where modulation is observed. This broadband strategy  
allows to  effectively remove the noise by fitting the data to the expected time dependent modulation (\ref{eq:annual}) or (\ref{eq:daily}).
As the next step one can use a resonance based cavity experiment to scan a single frequency bin where modulation is observed to pinpoint a precise value for the axion mass.

\section{Spectral properties in the rest frame}
\label{app:spectral properties in the rest frame}
In this appendix, we fulfill the technical details in Sec. \ref{subsec:axion-emission spectrum in laboratory frame}. In Eq. \eqref{eq:rho_rest(v_a)}, the $H_l(\tilde{p},\delta)$ is the partial wave expansion known as follows \cite{Liang:2018ecs}: 
\begin{widetext}
	\begin{subequations}
		\label{eq:3.2 a_solution ass}
		\begin{equation}
		\label{eq:3.2 a_solution ass_H}
		\begin{aligned}
		H_l(\tilde{p},\delta)
		&\equiv\sum_{n=0}^\infty\sum_{k=0}^n\frac{e^{-k\delta}}{2^n}
		\frac{(n+1)!}{k!(n-k)!}\frac{(-1)^k}{(k+1)^{l+3}}
		\Gamma(l+\frac{3}{2})f\left(\frac{1}{2}(l+3),\frac{1}{2}(l+4),
		l+\frac{3}{2};\frac{-(\tilde{p}/m_a)^2}{(k+1)^2}\right),
		\end{aligned}
		\end{equation}
		\begin{equation}
		\label{eq:3.2 a_solution ass_f}
		\begin{aligned}
		f\left(\frac{1}{2}(l+3),\frac{1}{2}(l+4),
		l+\frac{3}{2};\frac{-(\tilde{p}/m_a)^2}{(k+1)^2}\right)
		&\simeq\frac{1}{\Gamma(l+\frac{3}{2})}\left[
		1-\frac{(l+3)(l+4)}{4(k+1)^2}
		\frac{\Gamma(l+\frac{3}{2})}{\Gamma(l+\frac{5}{2})}(\tilde{p}/m_a)^2
		+{\cal O}(\tilde{p}/m_a)^4\right],  \\
		\end{aligned}
		\end{equation}
	\end{subequations}
\end{widetext}
where $f(a,b,c;z)$ is defined to be the regularized Gauss hypergeometric function ${}_2F_1(a,b,c;z)$, i.e. $f(a,b,c;z)\equiv\frac{1}{\Gamma(c)}{}_2F_1(a,b,c;z)$, see Refs. \cite{Abramowitz15:1972,Abramowitz6:1972} and recent article \cite{DiLella:2002ea}. We also note a useful fact in Eq. \eqref{eq:3.2 a_solution ass_f} that $f(a,b,c;z)$ has a simple quadratic behaviour in the non-relativistic limit $z\rightarrow0$.

Next, we turn to the normalization factor $N(\delta)$. According to Eq. \eqref{eq:3.2 a_solution ass}, the normalization factor obviously depends on $\delta$. We refer the detailed calculation to the work \cite{Liang:2018ecs}, while quoting results as follows
\begin{equation}
\label{eq:3.2 N}
N(\delta)=\left\{
\begin{aligned}
&0.434, &\delta=0.0  \\
&0.616, &\delta=0.5  \\
&0.798, &\delta=1.0
\end{aligned}
\right.
\end{equation}
In this work, we choose the intermediate value $\delta=0.5$ in the simulation, and we find the corresponding mean velocity of emitted axions is $\langle v_a\rangle\simeq0.6c$.

\section{Detailed calculation of strong gravitational lensing}
\label{app:detailed calculation of strong gravitational lensing}
It is shown in Refs. \cite{Patla:2013vza,Bertolucci:2017vgz} the signal of dark matter (DM) detection can be amplified as large as $\sim10^6$ by weak gravitational lensing for colinear stream of slow-moving particles. However, as shown in Sec. \ref{subsec:gravitational lensing} assumptions in Refs. \cite{Patla:2013vza,Bertolucci:2017vgz} do not apply to the case of AQNs in SHM. Unlike Refs. \cite{Patla:2013vza,Bertolucci:2017vgz} where the lensing is \textit{weak}, we need to calculate \textit{strong} gravitational deflection because there is a fixed $60\degree$ incident angle as shown in Fig. \ref{fig:assumption1}. Therefore, we will derive the amplification from first-principle Newtonian gravity. In what follows, we present a pedagogical approach from simple case to realistic situation.

\subsection{Simple case: weak gravitational lensing and coherent velocity}
\label{subapp:simple case}

\begin{figure}[h]
	\centering
	\captionsetup{justification=raggedright}
	\includegraphics[width=\linewidth]{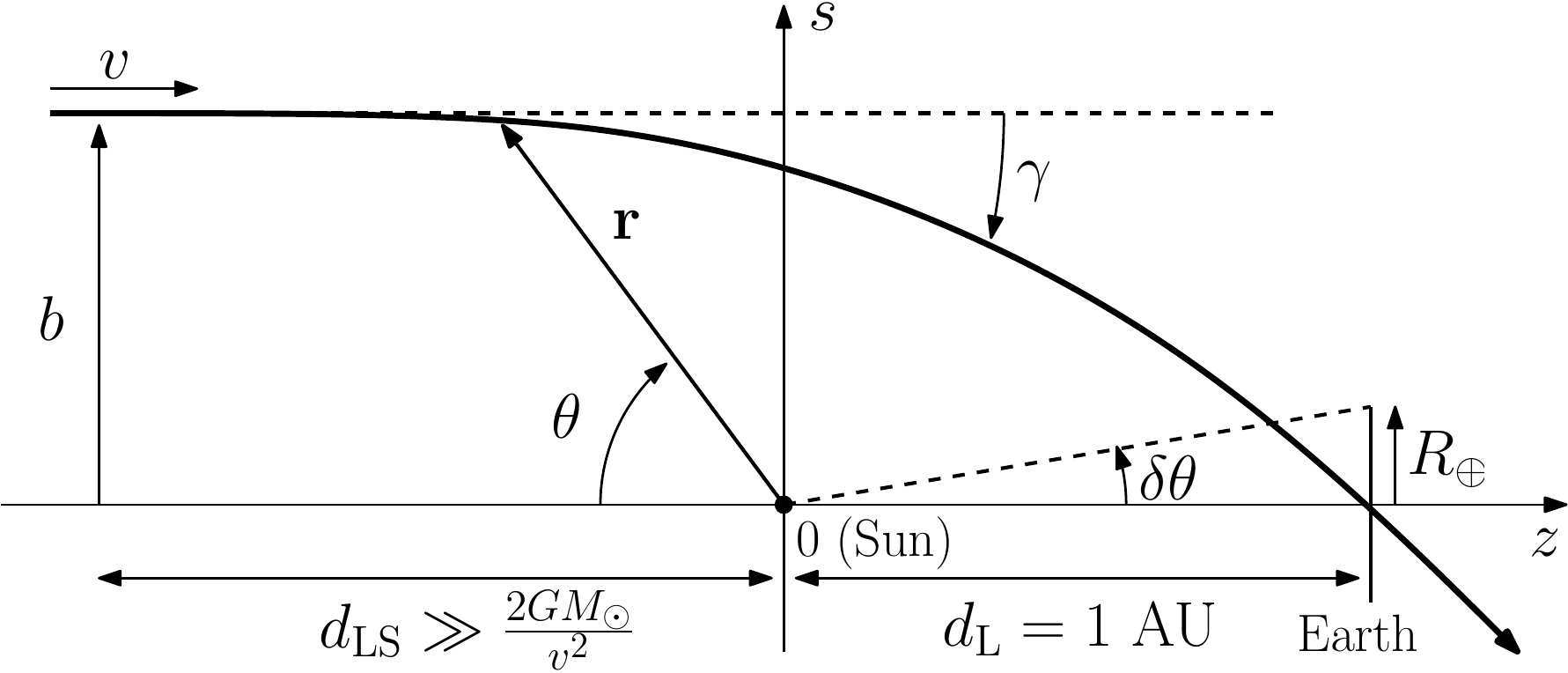}
	\caption{Weak lensing and coherent velocity: Here the flux comes from far distance $d_{\rm LS}\gg\frac{2GM_\odot}{v^2}$ with speed $v$ and impact parameter $b$. $\gamma$ is the deflection angle following its conventional definition. The Earth is approximated as 2D disk with radius $R_\oplus$. The cylindrical coordinate system is used.}
	\label{fig:coord weak lensing}
\end{figure}

We first analyze the simplest situation: a directed aligned flux with one and the same velocity $\mathbf{v}$, see Fig. \ref{fig:coord weak lensing}. Because of the good alignment, we expect the lensing is weak and the amplification should reach a similar agreement with Refs. \cite{Patla:2013vza,Bertolucci:2017vgz}.

Following the notation in Fig. \ref{fig:coord weak lensing}. we first write down the equations of motion from first-principle Newtonian gravity:
\begin{subequations}
	\label{eq:0.5mv2 and dot(theta)}
	\begin{equation}
	\label{eq:0.5mv2}
	\frac{1}{2}v^2
	=\frac{1}{2}\dot{r}^2+\frac{1}{2}\frac{l^2}{r^2}-\frac{GM_\odot}{r},
	\end{equation}
	\begin{equation}
	\label{eq:dot(theta)}
	\dot{\theta}
	=\dot{r}\frac{\rmd\theta}{\rmd r}=\frac{l}{r^2},\quad l=bv \ .
	\end{equation}
\end{subequations}
The set of differential equations describes a DM flux comes from a far distance $d_{\rm LS}$ with speed $v$ and impact parameter $b$. The Earth is approximated as 2D disk with radius $R_\oplus$ as we only concern the AQNs that have impact to the Earth. Solving for the differential equations, we have
\begin{subequations}
\label{eq:dot(r) and d theta/dr}
\begin{equation}
\label{eq:dot(r)}
\dot{r}=\pm\sqrt{v^2+\frac{2GM_\odot}{r}-\frac{l^2}{r^2}}\ ,
\end{equation}
\begin{equation}
\label{eq:d theta/dr}
\frac{\rmd\theta}{\rmd r}
=\frac{1}{\dot{r}}\frac{l}{r^2}
=\pm\frac{l}{r^2}
\left(\sqrt{v^2+\frac{2GM_\odot}{r}-\frac{l^2}{r^2}}\right)^{-\frac{1}{2}}\ ,
\end{equation}
\end{subequations}
where the sign $\pm$ depends on the time before/after sthe distance of closest approach $r_{\rm min}$ defined as:
\begin{equation}
\label{eq:r_min}
r_{\rm min}
=\frac{GM_\odot}{v^2}\left[
\sqrt{1+\left(\frac{lv}{GM_\odot}\right)^2}-1
\right]\ ,
\end{equation}
Now, denoting a set of dimensionless parameters for convenience:
\begin{subequations}
	\label{eq:w, a, and w_min}
	\begin{equation}
	\label{eq:w}
	w\equiv\frac{v}{l}r=\frac{r}{b}\ ,
	\end{equation}
	\begin{equation}
	\label{eq:a}
	a\equiv\frac{2GM_\odot}{vl}=\frac{2GM_\odot}{v^2b}\ ,
	\end{equation}
	\begin{equation}
	\label{eq:w_min}
	w_{\rm min}\equiv\frac{r_{\rm min}}{b}
	=\frac{1}{2}a\left(\sqrt{1+\left(\frac{2}{a}\right)^2}-1\right)\ .
	\end{equation}
\end{subequations}
Integrating out Eq. \eqref{eq:d theta/dr}, we obtain an explicit equation of $\theta$ as a function of $r$:
\begin{widetext}
\begin{equation}
\label{eq:theta}
\theta(r)
=\left\{
\begin{aligned}
&\tan^{-1}\left(\frac{a}{2}\right)
-\tan^{-1}\left(
\frac{-1+\frac{1}{2}aw}{\sqrt{-1+w(a+w)}}
\right),
&t<t(r_{\rm min}),\\
&\tan^{-1}\left(\frac{a}{2}\right)+\pi
+\tan^{-1}\left(
\frac{-1+\frac{1}{2}aw}{\sqrt{-1+w(a+w)}}
\right),
&t\geq t(r_{\rm min}). 
\end{aligned}
\right.
\end{equation}
\end{widetext}

As a quick check, note the deflection angle $\gamma$:
\begin{equation}
\label{eq:gamma}
\gamma
\equiv\left.\theta\right|_{t\rightarrow\infty}
-\left.\theta\right|_{t\rightarrow-\infty}-\pi  
=2\tan^{-1}\left(\frac{GM_\odot}{v^2b}\right)\ ,
\end{equation}
which recovers to classic formula of (Newtonian) deflection angle in the weak lensing limit $b\gg\frac{GM_\odot}{v^2}$.

Now, we want to determine condition such that the flux will hit the Earth, that is
\begin{equation}
\label{eq:theta(d_L)-pi}
|\theta(d_{\rm L})-\pi|
\leq\delta\theta,\quad
\delta\theta\equiv\frac{R_\oplus}{d_{\rm L}}
=4.263\times10^{-5}\left(\frac{1{\rm AU}}{d_{\rm L}}\right)\ .
\end{equation}
Perturbatively solving the inequality \eqref{eq:theta(d_L)-pi}, we conclude to hit the Earth the impact parameter $b$ must be constraint within a small ring $(b\pm\delta b)$:
\begin{equation}
\label{eq:b and delta b}
b(v)=\sqrt{\frac{2GM_\odot d_{\rm L}}{v^2}},\quad
\delta b = R_\oplus\ .
\end{equation}
From Eqs. \eqref{eq:b and delta b}, the amplification factor $A_0(v)$ is just the ratio of the incident ring to the cross section without gravity:
\begin{equation}
\label{eq:A_0(v)}
\begin{aligned}
A_0(v)
&=\frac{2\pi b(v)\cdot 2\delta b_{\rm max}}{\pi R_\oplus^2}
=\sqrt{\frac{32GM_\odot d_{\rm L}}{v^2R_\oplus^2}} \\
&=4494
\left(\frac{d_{\rm L}}{1{\rm AU}}\right)^{\frac{1}{2}}
\left(\frac{220\kmps}{v}\right)\ .
\end{aligned}
\end{equation}
Therefore, we conclude the amplification (weak lensing and coherent velocity) is of order $\sim10^3-10^4$ for a typical galactic speed $220\kmps$. We conclude we reach a similar, but not the same, magnification as calculated in Refs. \cite{Patla:2013vza,Bertolucci:2017vgz}. But we warn the readers such comparison for illustrative purpose only as the assumptions in this work are different from Refs. \cite{Patla:2013vza,Bertolucci:2017vgz}, as explained in the Sec. \ref{subsec:gravitational lensing}. 

Lastly, we comment on one subtle point regarding to the validity of equations \eqref{eq:0.5mv2 and dot(theta)}. That is, the initial position of DM flux $d_{\rm LS}$ to be ``sufficiently'' far distance such that we can safely assume the initial velocity $v$ is not affected by gravity. The condition of validity can be expressed as:
\begin{equation}
\label{eq:d_LSs}
\frac{1}{2}v^2
\gg\frac{GM_\odot}{d_{\rm LS}}\ ,
\end{equation}
alternatively we can define a dimensionless parameter $\varepsilon$ as:
\begin{equation}
\label{eq:varepsilon}
\varepsilon
\equiv\frac{2GM_\odot}{v^2d_{\rm LS}}
\ll1\ .
\end{equation}
Although it sounds to be a trivial point, the parameter $\varepsilon$ turns out to be an important parameter in the next subsection. As we will see, the amplification derived in the realist case will be expressed in terms of $\varepsilon$.

\subsection{Realistic case: strong gravitational deflection and dispersive velocity}
\label{subapp:realistic case}
\begin{figure}[h]
	\centering
	\captionsetup{justification=raggedright}
	\includegraphics[width=\linewidth]{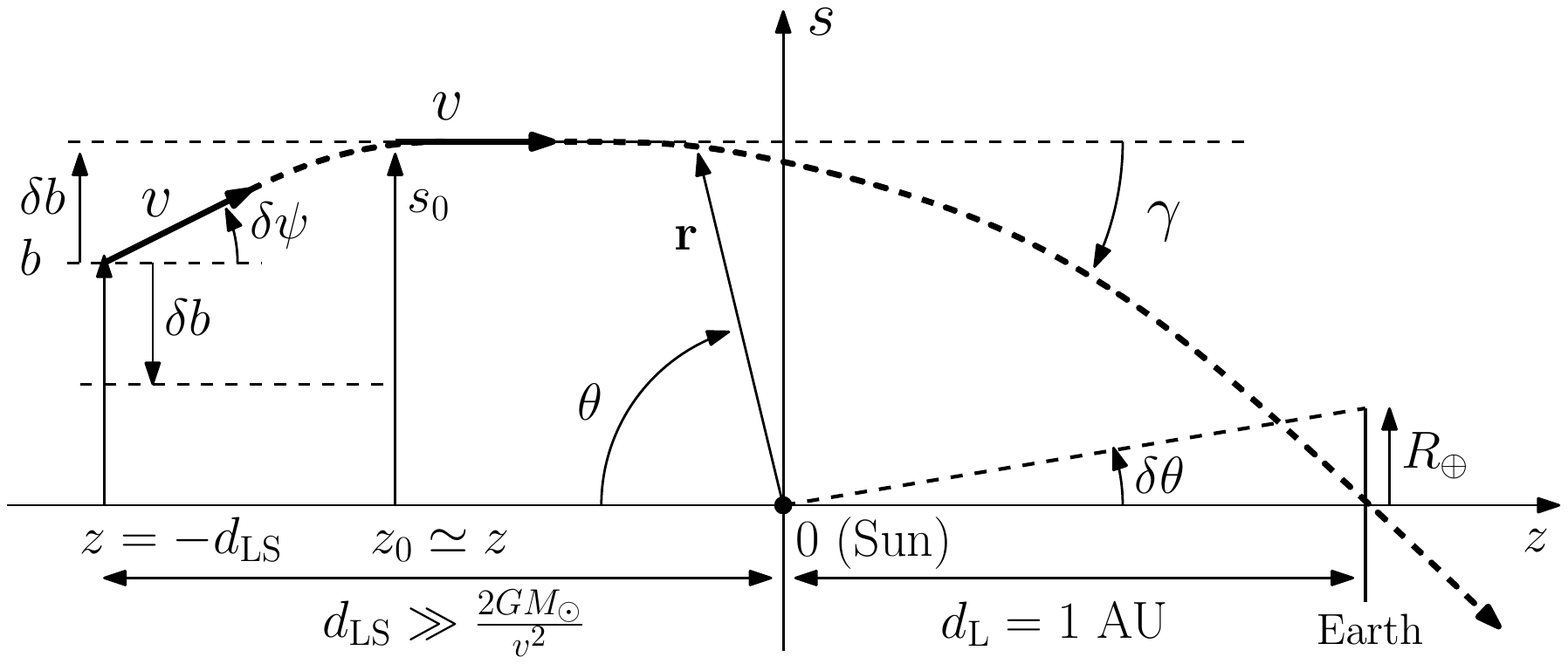}
	\caption{Strong deflection and dispersive velocity: Here the flux comes from far distance $d_{\rm LS}$ with impact parameter $b$. Its has initial speed $v$ with a small tilted angle $\delta\psi$ due to presence of dispersion. After traveling for a short distance (compared to $d_{\rm LS}$), it returns to a parallel alignment as in the simple case Fig. \ref{fig:coord weak lensing}. We require the `new' impact parameter $s_0$ is within the focusing ring $(b\pm\delta b)$ in order to hit the Earth. The Earth is approximated as 2D disk with radius $R_\oplus$. The cylindrical coordinate system is used.}
	\label{fig:coord strong deflection}
\end{figure}

In reality the AQN flux is obviously dispersive and the gravitational deflection is large because of the fixed $60\degree$ incident angle, see Fig. \ref{fig:assumption1}. To justify this realistic case, we first consider the configuration as shown in Fig. \ref{fig:coord strong deflection}. We assume initially there is a small tilted angle $\delta\psi$. We ask how big $\delta\psi$ is allowed? To answer this question, we trace the particle trajectory after its launch for some distance. Due to the existence the gravitation, the particle goes back to the parallel aligned case at $(z_0,s_0)$. Therefore, the only requirement is
\begin{equation}
\label{eq:s_0-b}
|s_0-b|\leq\delta b\ ,
\end{equation}
such that the particle can still hit the Earth.

Now we want to find $(z_0,s_0)$ as a function of $\delta\psi$. We modify the equation of motion in Eqs. \eqref{eq:0.5mv2 and dot(theta)} as
\begin{equation}
\label{eq:dv(t) dt}
\frac{\rmd}{\rmd t}\mathbf{v}
=-\hat{\mathbf{r}}\frac{GM_\odot}{r^2}
\simeq-\hat{\mathbf{r}}\frac{GM_\odot}{z^2}\ .
\end{equation}
In the last step, we use the fact $|z|\simeq d_{\rm LS}\gg s$. We present only the radial component  $s$, which is of interest
\begin{equation}
\label{eq:dv_s dt}
\frac{\rmd}{\rmd t}v_s
= v_s\frac{\rmd v_s}{\rmd s}
\simeq-\frac{GM_\odot}{|z|^3}s
\simeq-\frac{GM_\odot}{d_{\rm LS}^3}s\ .
\end{equation}
Solving this equation, we obtain
\begin{equation}
\label{eq:v_s2-v_s02}
v_s^2-v_{s,0}^2
\simeq v^2\delta\psi
=\frac{GM_\odot}{d_{\rm LS}^3}(s_0^2-s^2)
\simeq\frac{GM_\odot}{d_{\rm LS}^3} 2b(v) \Delta s\ ,
\end{equation}
where we define $\Delta s\equiv s_0-b$ and utilize the fact $v_{s,0}=0$ and $v_s=v\sin\delta\psi\simeq v\delta\psi$ for small $\delta\psi$ in the second step, and $s_0=b(v)+\mathcal{O}(\Delta s)$ in the last receptively. To ensure the particle still hit the Earth, we require $\Delta s\leq\delta b$. Therefore using Eqs. \eqref{eq:b and delta b} and \eqref{eq:v_s2-v_s02}, we have the following inequality:
\begin{equation}
\label{eq:delta psi}
\delta\psi
\leq\frac{v^3R_\oplus d_{\rm L}^{1/2}}{(2GM_\odot)^{3/2}}\varepsilon^3\ ,
\end{equation}
where $\varepsilon\ll1$ is a convenient parameter defined in Eq. \eqref{eq:varepsilon}.

From the flux distribution \eqref{eq:d dot(N) dv}, we now define the angular flux spectrum as follows:
\begin{equation}
\label{eq:Phi(theta,v)}
\begin{aligned}
\Phi(\theta,v)
&\equiv\frac{\rmd^2\dot{N}}{\rmd v\rmd(\cos\theta)}  \\
&=C\int\rmd\psi\int\rmd\varphi\, v^3 e^{-\frac{v^2}{2\sigma^2}}\sin\psi\cos\psi\times \\
&\quad\times\exp\left[
-\frac{v\mu}{\sigma^2}(\cos\psi\cos\theta-\sin\psi\cos\varphi\sin\theta)
\right]\ .
\end{aligned}
\end{equation}
Then a more realistic estimate of enhancement is therefore
\begin{widetext}
\begin{equation}
\label{eq:A(theta,v)}
\begin{aligned}
A(\theta,v)
&=\frac{\Phi(\theta,v)}{\Phi(0,v)}A_0(v)
=\frac{
	\int_{\pi-\delta\psi}^\pi d\psi\int_0^{2\pi}d\varphi~
	\sin\psi\cos\psi
	\exp\left[{-\frac{v\mu}{\sigma^2}
		(\cos\psi\cos\theta-\sin\psi\cos\varphi\sin\theta)}
	\right]
}{
	\int_\frac{\pi}{2}^\pi d\psi\int_0^{2\pi}d\varphi~
	\sin\psi\cos\psi
	\exp\left[{-\frac{v\mu}{\sigma^2}
		\cos\psi}
	\right]
}
\sqrt{\frac{32GM_\odot d_{\rm L}}{v^2R_\oplus^2}}  \\
&\simeq\left(\frac{v\mu}{\sigma^2}\right)^2
\frac{
	\delta\psi^2\exp(\frac{v\mu}{\sigma^2}\cos\theta)
}{
	1+(\frac{v\mu}{\sigma^2}-1)\exp(\frac{v\mu}{\sigma^2})
}
\sqrt{\frac{32GM_\odot d_{\rm L}}{v^2R_\oplus^2}}
\, [1+{\cal O}(\delta\psi)]\ .
\end{aligned}
\end{equation}
\end{widetext}
Substituting inequality \eqref{eq:delta psi} insto Eq. \eqref{eq:A(theta,v)}, we arrive
\begin{equation}
\label{eq:A(theta,v) leq}
A(\theta,v)
\leq\varepsilon^6\left(\frac{v\mu}{\sigma^2}\right)^2
\frac{
	4v^5\exp(\frac{v\mu}{\sigma^2}\cos\theta)
}{
	1+(\frac{v\mu}{\sigma^2}-1)\exp(\frac{v\mu}{\sigma^2})
}
\frac{R_\oplus d_L^{3/2}}{(2GM_\odot)^{5/2}}\ .
\end{equation}
The condition is clearly sensitive to the value of $\varepsilon$. Subjecting to the constraint \eqref{eq:varepsilon}, $\varepsilon$ has to be small. In Fig. \ref{fig:A_max} we choose the marginal value $\varepsilon=\frac{1}{3}$, namely the gravitational effect does not contribute to the initial total energy more than 25\%. We note that for $\varepsilon$ greater than this value, the flux spectrum \eqref{eq:Phi(theta,v)} may fail because strong gravitation becomes comparable to initial kinetic energy. Finally, we also comment that our estimation agrees with Ref. \cite{Lee:2013wza} although different assumptions and models are adopted.

\section{On sensitivity of the main results to the AQN's parameters}
\label{app:on sensitivity of the main results to the AQN's parameters}
The main goal of this appendix is argue that thee main resultss of this work are not very sensitive to the parameters of the model, such as size distribution of AQN (paramters $\alpha$ and $B_{\rm min}$) and flux distribution of incoming AQNs (e.g. the annual/daily modulation and solar gravitation). 

Before presenting the simulation results, we describe the parameters considered in the work as summarized in Table \ref{tab:summary of some results}. First, we choose the set of $(\alpha,B_{\rm min})$ in Table \ref{tab:mean B} such that $\langle B\rangle\gtrsim10^{25}$, subjecting to the observational constraints by IceCube and and ANITA as discussed in Sec. \ref{sec:the AQN model and its axion emission mechanism}. Next, we consider two extreme cases of annual modulation by modifying the mean galactic velocity $\mu$ by adding (subtracting) the orbital speed of the Earth. In addition, we also consider the gravitation of solar system because it implies an additional escape velocity $\sim42.1\kmps$ at Earth distance. To take into account of such effect, we add an additional magnitude to the initial speed for each AQN by $\sim42.1\kmps$ in the simulation based on the flux distribution \eqref{eq:d dot(N) dv}.

\begin{figure*}[!htp]%[!htp]
	\centering
	\captionsetup{justification=raggedright}
	\begin{subfigure}[b]{0.47\linewidth}
	\includegraphics[width=\linewidth]{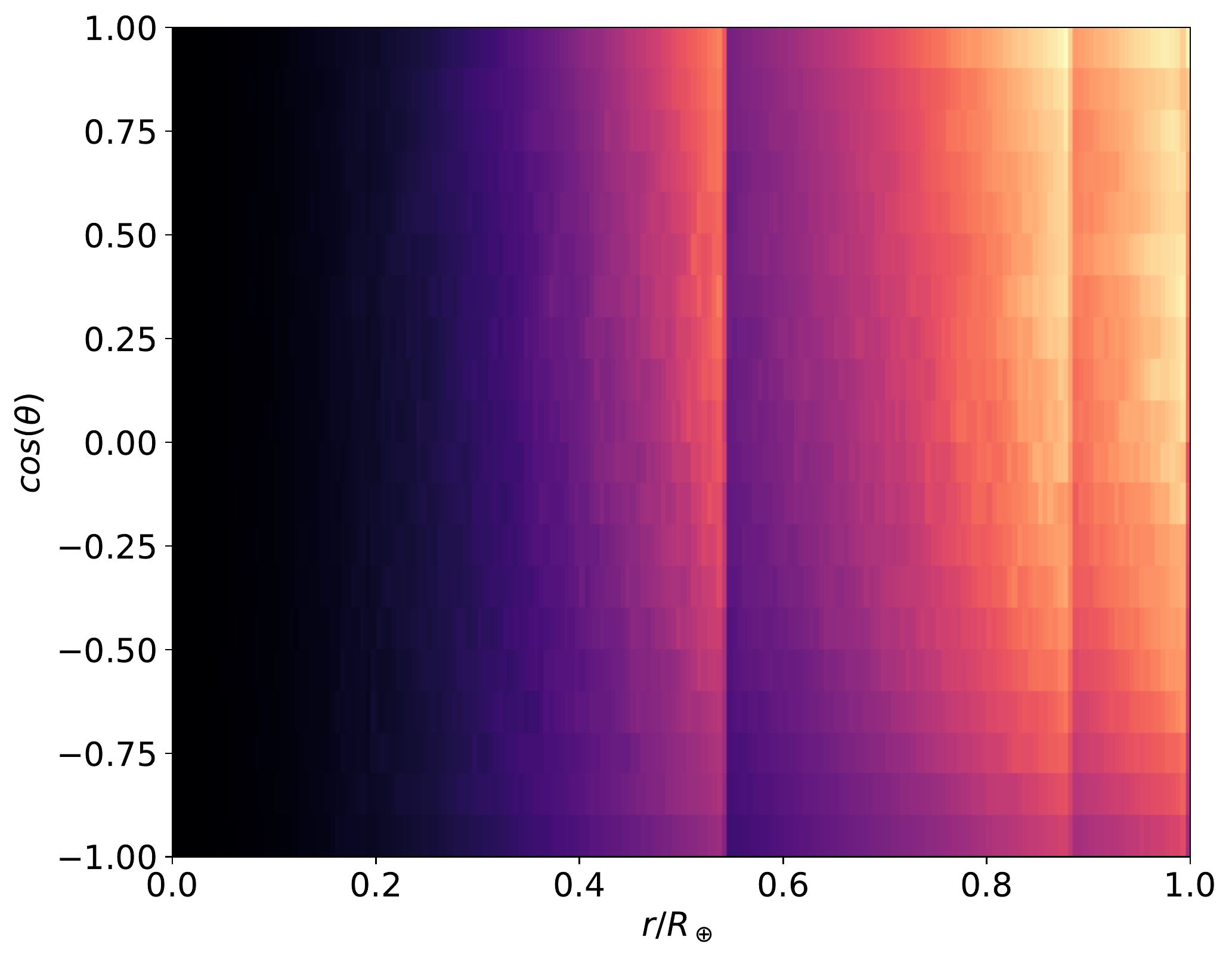}
	% \caption{$\alpha=2.5$, $\mu=V_\odot+V_\oplus$}
	\caption{}
	\label{fig:q_alpha25_muPlus}
	\end{subfigure}
	\begin{subfigure}[b]{0.47\linewidth}
	\includegraphics[width=\linewidth]{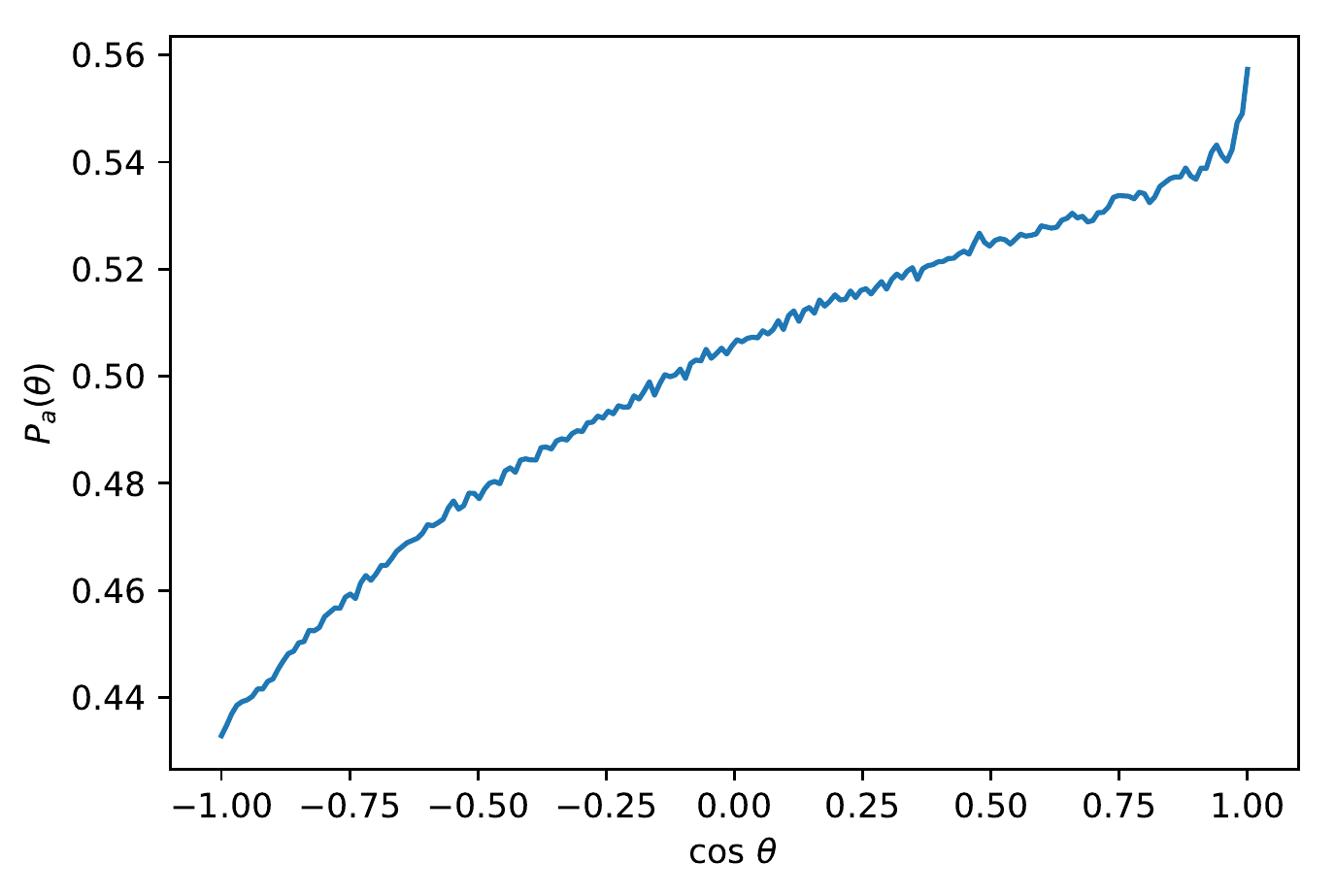}
	% \caption{$\alpha=2.5$, $\mu=V_\odot+V_\oplus$}
	\caption{}
	\label{fig:Pa_alpha25_muPlus}
	\end{subfigure}
	\begin{subfigure}[b]{0.47\linewidth}
	\includegraphics[width=\linewidth]{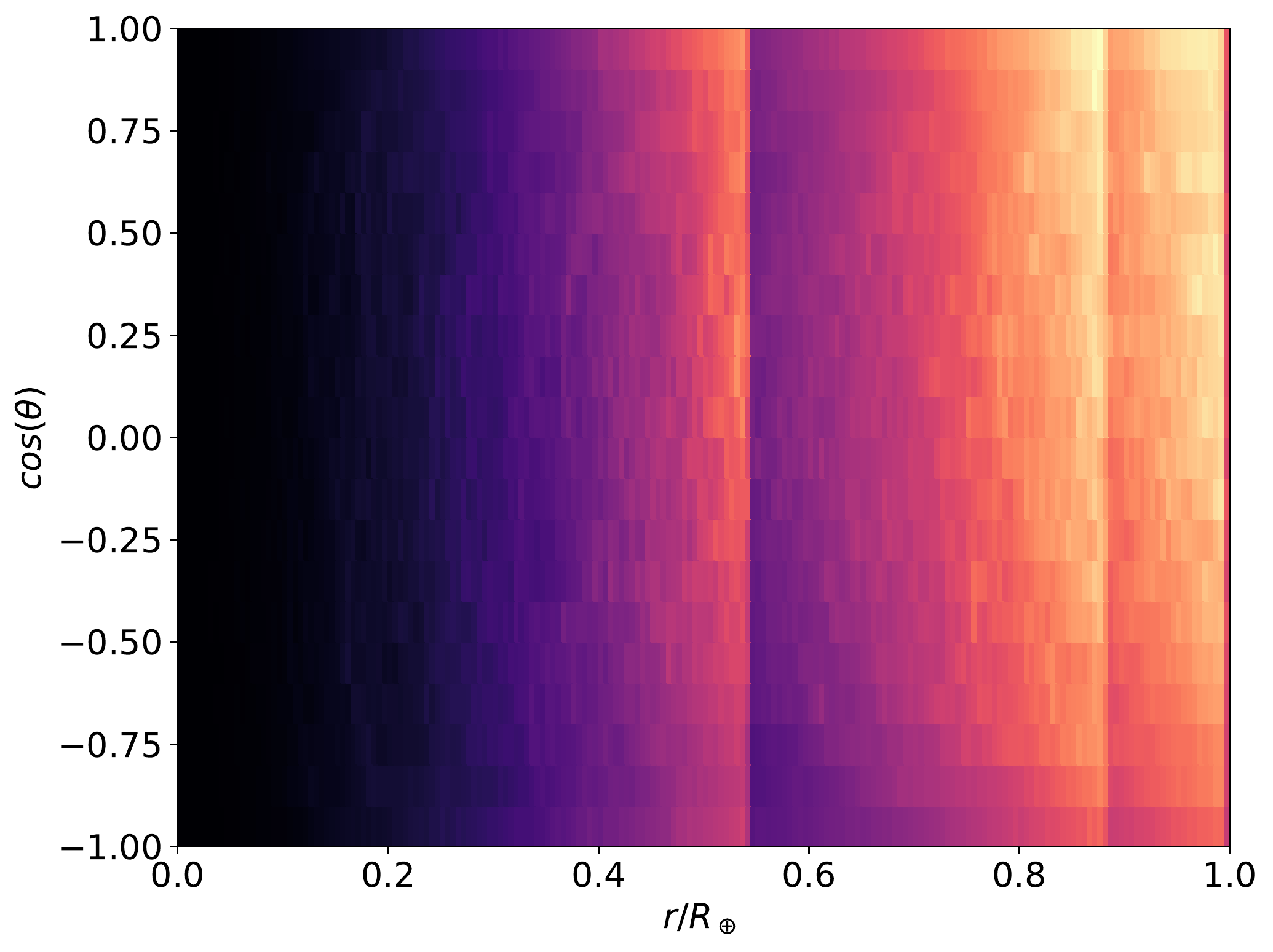}
	% \caption{$\alpha=2.0$}
	\caption{}
	\label{fig:q_alpha20}
	\end{subfigure}
	\begin{subfigure}[b]{0.47\linewidth}
	\includegraphics[width=\linewidth]{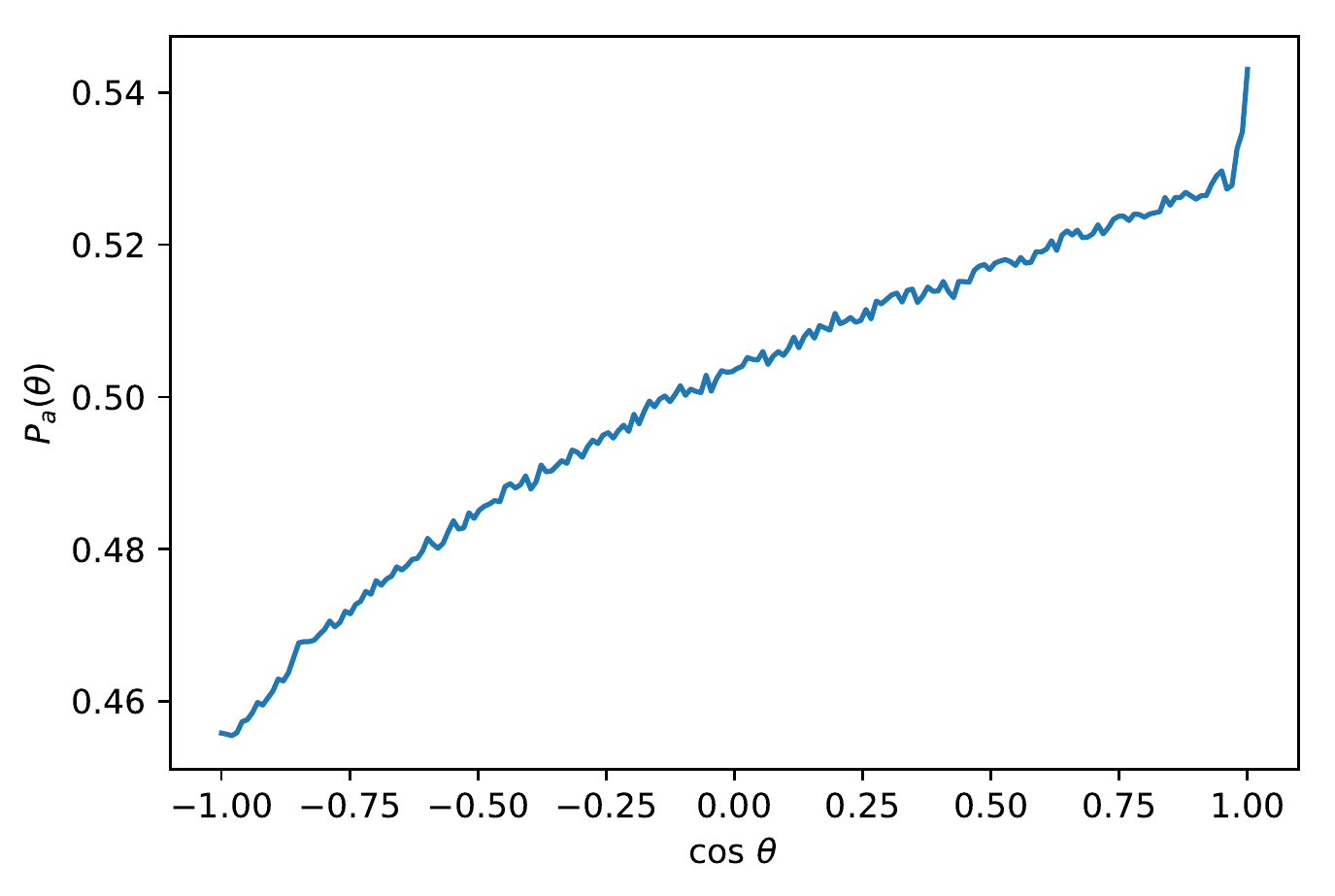}
	% \caption{$\alpha=2.0$}
	\caption{}
	\label{fig:Pa_alpha20}
	\end{subfigure}
	\begin{subfigure}[b]{0.47\linewidth}
	\includegraphics[width=\linewidth]{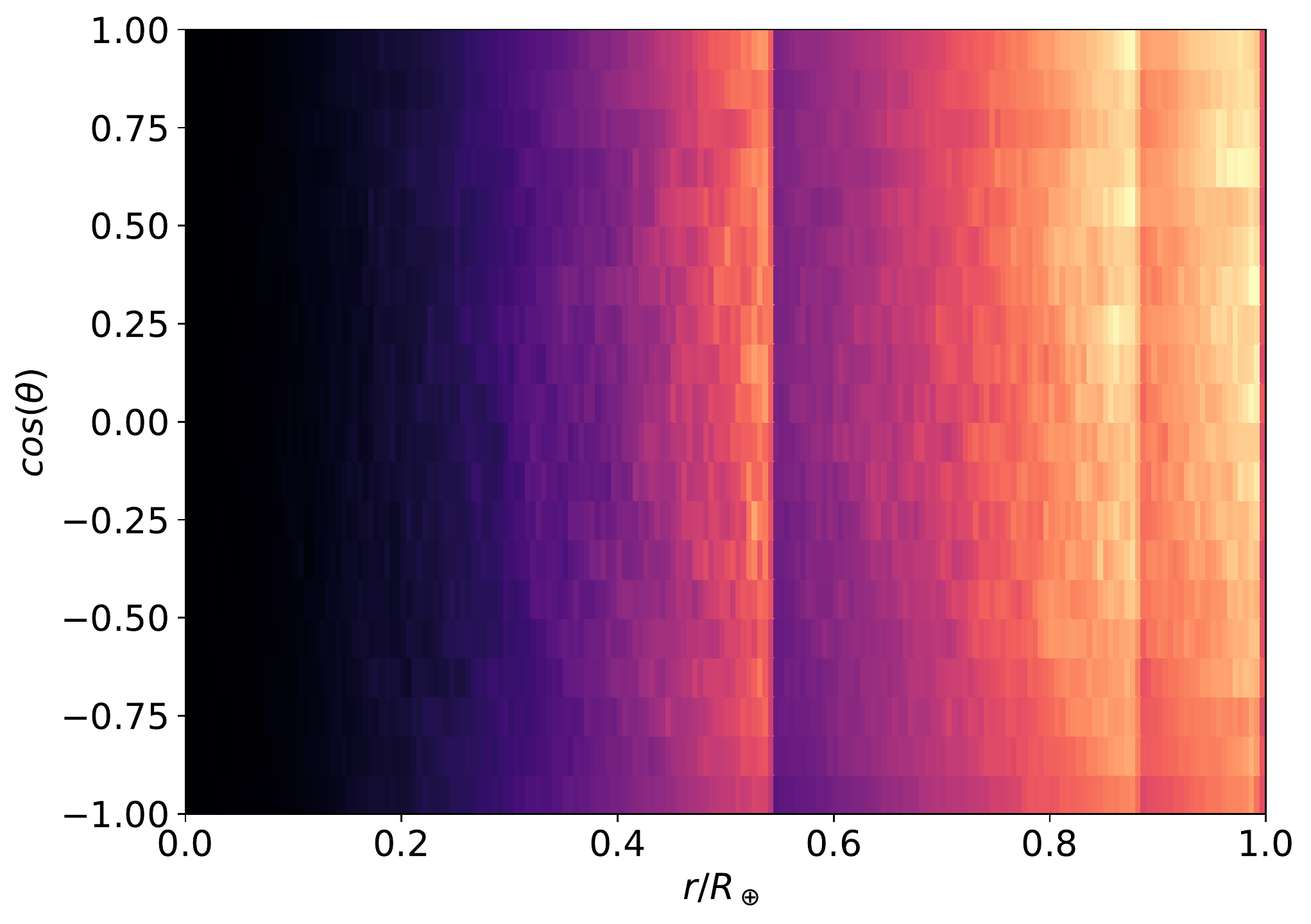}
	% \caption{$\alpha=(1.2,2.5)$, $B_{\rm min}=10^{23}$}
	\caption{}
	\label{fig:q_alphaPW_Bmin1e23}
	\end{subfigure}
	\begin{subfigure}[b]{0.47\linewidth}
	\includegraphics[width=\linewidth]{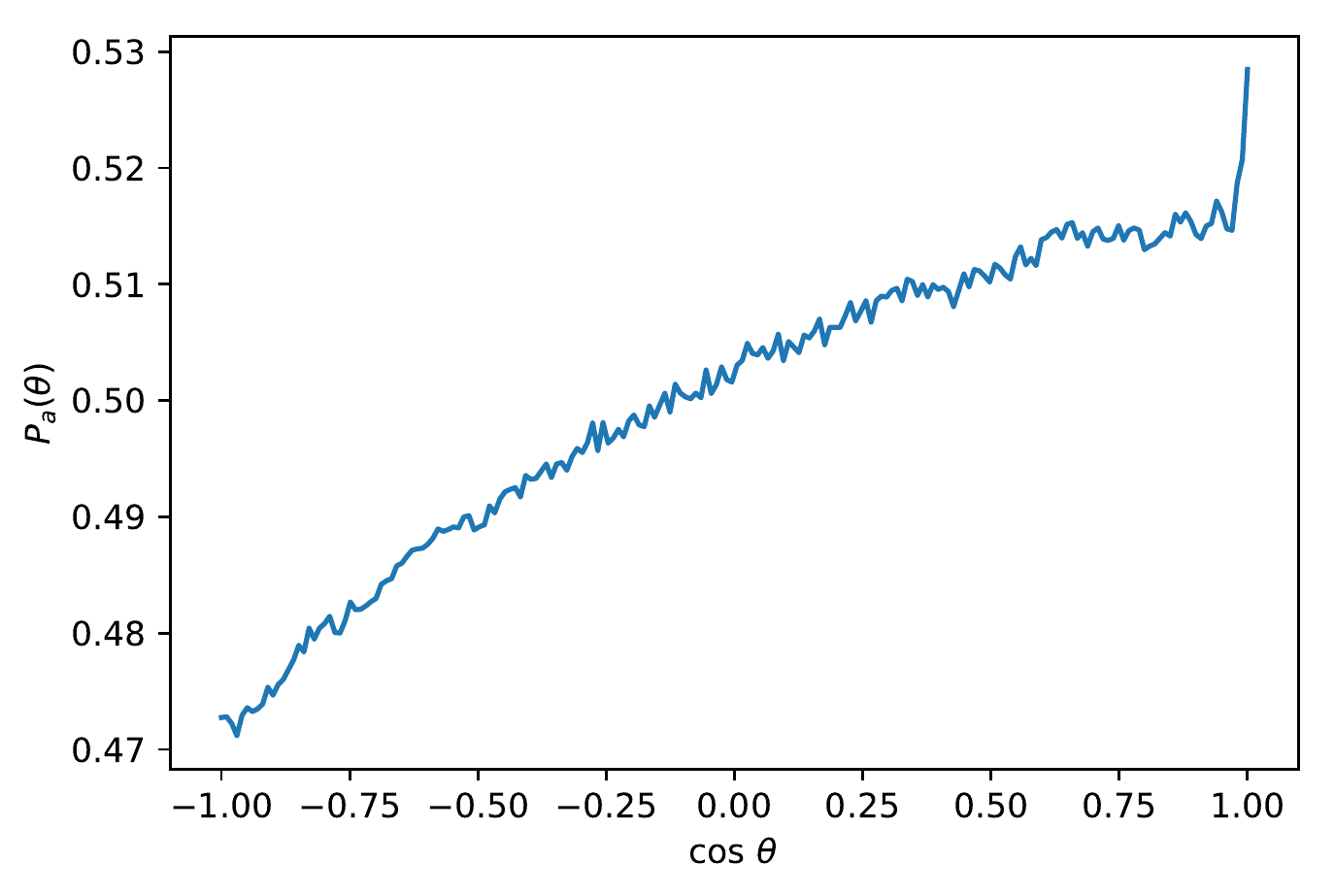}
	% \caption{$\alpha=(1.2,2.5)$, $B_{\rm min}=10^{23}$}
	\caption{}
	\label{fig:Pa_alphaPW_Bmin1e23}
	\end{subfigure}
	\caption{Summary of simulations. The heat emission profiles $q(r,\theta)$ are presented in the left column ($2\times10^5$ examples), following by the azimuthal distribution $P_a(\theta)$ ($10^8$ samples) on the right. 
	Each row corresponds to difference choices of parameters ($B_{\rm min}=3\times10^{24}$ unless specified): $\alpha=2.5$, $\mu=V_\odot+V_\oplus$ (Figs. \ref{fig:q_alpha25_muPlus}, \ref{fig:Pa_alpha25_muPlus}); $\alpha=2.0$ (Figs. \ref{fig:q_alpha20}, \ref{fig:Pa_alpha20}); $\alpha=(1.2,2.5)$, $B_{\rm min}=10^{23}$ (Figs. \ref{fig:q_alphaPW_Bmin1e23}, \ref{fig:Pa_alphaPW_Bmin1e23}).
	All choices of parameters in Table \ref{tab:summary of some results} share similar features.}
	\label{fig:other plots}
\end{figure*}

As mentioned earlier, we find the simulation results are largely similar in all cases. For the sake of brevity, we only present one case of annual modulation [$(\alpha,B_{\rm min})=(2.0,3\times10^{24})$, and $\mu=V_\odot+V_\oplus$], and two cases with different choices of $(\alpha,B_{\rm min})$: $(2.0,3\times10^{24})$ and $({\rm piesewise},10^{23})$ as shown in Figs. \ref{fig:other plots}. Apparently, the heat emission profiles $q(r,\theta)$ and the azimuthal distributions $P_a(\theta)$ share similar features as presented in Figs. \ref{fig:q(r,theta)_alpha25} and \ref{fig:P_a(r,theta)_alpha25}. We conclude the main results are not senstive to the parameters of model.

\FloatBarrier
 
\bibliography{modulation}

\end{document}